\documentclass[a4paper,11pt]{article}
\usepackage{jheppub} 
\usepackage{lineno}
\usepackage[dvipsnames]{xcolor}
\usepackage{slashed}

\title{\boldmath A note on partially massless supergravity}

\author[a]{Nicolas Boulanger,}
\author[a]{Guillaume Lhost,}
\author[a,1]{Sylvain Thomée\note{Research Fellow of the F.R.S.-FNRS (Belgium)}}
\affiliation[a]{Service de Physique de l’Univers, Champs et Gravitation,\\Université de Mons -- UMONS,\\20 place du Parc, 7000 Mons, Belgium}

\emailAdd{sylvain.thomee@umons.ac.be}

\abstract{
We analyse the couplings of a partially massless 
spin-2 field with a doublet of massless, real spin-3/2 fields.
In the flat limit, this spectrum coincides with the spectrum of  
${\cal N}=2$ pure supergravity around anti-de Sitter spacetime AdS$_4$.
We classify all the possible parity-invariant, non-Abelian 
deformations of the free theory that lead to a deformation of the Lagrangian.
By doing this, we re-derive a non-Abelian vertex recently found 
in \href{https://inspirehep.net/literature/2856848}{2412.04982 [hep-th]} 
by Yu. M. Zinoviev following a different approach, 
and show that the corresponding non-Abelian gauge algebra closely resembles 
the one of ${\cal N}=2$ supergravity around AdS$_4$.
The gauge-algebra deformation is however obstructed, at next order. 
Then, we add a massless vector field together with a massive 
spin-3/2 field, and find two nontrivial vertices mixing these fields
with a single massless gravitino. Still, the gauge algebra remains 
obstructed at next order, therefore excluding the possibility to 
make local the global supersymmetry algebra found in recent works on 
partially massless supermultiplets in AdS$_4$. Finally, we argue that 
the two problems encountered are simultaneously solved by the addition 
of the masseless graviton, leading to ${\cal N}=1$ pure conformal 
supergravity around AdS$_4$ as the only consistent theory coupling 
partially massless spin-2 fields to massless and massive spin-3/2 fields. 
}

\begin{document}
\maketitle
\flushbottom

\section{Introduction}\label{sec:Introduction}

The partially massless (PM) spin-2 field 
\cite{Deser:1983tm,Deser:1983mm,Higuchi:1986py,Higuchi:1986wu,Deser:2001pe,Deser:2001us,Deser:2001wx}
is of special interest in the cosmological context, since it propagates and 
is unitary only on the de Sitter (dS$_4$) background relevant to the inflationary
epoch. Since the mass of the PM spin-2 field saturates the Higuchi bound 
\cite{Higuchi:1986py,Deser:2001wx}, the PM spin-2 field, if 
it existed in our early Universe, would have behaved like a light field 
during inflation, therefore possibly leaving an imprint on the cosmic 
microwave background
\cite{Baumann:2017jvh,Franciolini:2017ktv,Goon:2018fyu}.
The mass of the PM spin-2 field is protected by gauge 
invariance and is uniquely fixed in terms of the cosmological constant.
In fact, the present epoch of our Universe can also be approximated 
by a de Sitter phase, with its accelerated expansion \cite{SupernovaCosmologyProject:1998vns, SupernovaSearchTeam:1998fmf}. 
Due to the smallness of the observed positive cosmological constant \cite{Planck:2018vyg}, 
the mass for the PM spin-2 field is below the upper-bound on the mass 
of the graviton deduced from the gravitational waves detections \cite{LIGOScientific:2018dkp, LIGOScientific:2020tif, LIGOScientific:2021sio}.

Although the partially massless spin-2 field only propagates on (anti-) 
de Sitter (A)dS$_4$ background, its Stueckelberg off-shell description 
admits a flat limit \cite{Zinoviev:2008ze,Boulanger:2018shp} that shows 
that a partially massless spin-2 field propagates the modes of 
helicity $\pm 2$ for a massless graviton, plus the modes of helicity $\pm 1$ 
for a massless vector field, in a way that provides an alternative to the 
original Hamiltonian analysis of \cite{Deser:2001us}.

There have been many studies on the self-couplings of a PM spin-2 field, 
see \cite{Boulanger:2019zic} and references therein, 
but relatively few analyses of the couplings of the PM spin-2 field with other 
types of fields, with the notable exceptions of 
\cite{Joung:2014aba,Joung:2019wwf}
where the gravitational coupling of the PM spin-2 field was investigated, 
together with the interactions including massless vectors \cite{Boulanger:2024hrb}. 
In the case of the gravitational coupling of the PM spin-2 field, 
the latter work strengthens the no-go results of 
\cite{Joung:2019wwf} and precludes the possibility that the 
non-geometrical coupling considered in \cite{Joung:2019wwf}
could be pushed to higher orders in interactions.
It was also shown in \cite{Boulanger:2024hrb} that the vectors 
couple to the massless and partially massless spin-2 fields 
in a way that reproduces several copies of conformal gravity coupled 
to Yang-Mills sectors.

In a previous work \cite{Boulanger:2023lgd}, 
we considered the interactions of the partially massless spin-2 field 
with a massive spin-3/2 field,
that together form an on-shell spectrum of particles with helicities 
$(\pm 2, \pm 1, \pm 3/2, \pm 1/2)$, thereby (naively) offering of possibility 
for supersymmetry.
Two consistent couplings were discovered in  \cite{Boulanger:2023lgd}, 
giving a hope that it might indeed be possible to find a locally supersymmetric theory 
with a PM spin-2 field in the spectrum. One of the couplings in 
\cite{Boulanger:2023lgd} is very similar to the minimal coupling of a massive 
spin-3/2 field to gravity, except that now the graviton is partially massless. 
Actually, this spectrum of fields is too small to carry the action 
of rigid supersymmetry, as it was shown in \cite{Garcia-Saenz:2018wnw} 
that the shortest partially massless supermultiplet necessitates 
the adjunction of a pair of massless fields with helicities $(\pm 3/2 , \pm 1)\,$ 
to the PM spin-2 and the massive spin-3/2 fields in AdS$_4$. 
A natural question that we address in this article is the possibility for 
couplings among the four types of fields that constitute the shortest 
partially massless supermultiplet of \cite{Garcia-Saenz:2018wnw}.
In particular, in this work we investigate whether interactions among 
those fields can make local the global rigid supersymmetry carried by 
that multiplet. As we show in this article, the result is negative, 
unless one adds \emph{two} extra gauge fields to the spectrum: 
the massless graviton plus another massless gravitino. 
Using the analyses of \cite{Fradkin:1985am,Bobev:2021oku},
we find that the resulting spectrum of fields exactly 
coincides with the 
spectrum of ${\cal N}=1$ pure conformal supergravity around AdS$_4$.

Similarly to the observation that conformal gravity is the 
only consistent way to couple the PM spin-2 field to gravity 
\cite{Joung:2014aba,Joung:2019wwf,Boulanger:2024hrb}, it is 
therefore natural to expect that conformal supergravity 
(see \cite{Fradkin:1985am} for a review) 
should give the only non-Abelian interactions of a PM spin-2 field 
with massless spin-3/2 fields around AdS$_4$.
The spectrum that consists of a PM spin-2 field and two 
massless gravitini is exactly the same as the one of pure 
${\cal N}=2$ supergravity around AdS$_4$, 
supporting the possibility that there might be consistent 
non-Abelian interactions among those fields, viewed as a sub-sector of 
conformal supergravity. 

To summarize, the goal of this paper is twofold: we investigate whether it is 
possible to make local the global (i.e., rigid) supersymmetry 
of \cite{Garcia-Saenz:2018wnw}, and we want to determine whether the only 
consistent interactions of a PM spin-2 field with a doublet of massless 
spin-3/2 fields around AdS$_4$ is conformal supergravity, 
so as to provide the supersymmetric extension of the results in 
\cite{Joung:2014aba,Joung:2019wwf,Boulanger:2024hrb}.
We show that the answer to both questions is positive, and present arguments 
that indicate that pure ${\cal N}=1$ conformal supergravity solves 
both problems simultaneously. 

More in detail, in this work 
we first consider a free theory in AdS$_4$ containing a partially 
massless spin-$2$ field, a massive spin-$3/2$ field, a massless spin-$3/2$ 
field and a massless spin-$1$ field 
which enjoys a rigid $\mathcal{N}=1$ supersymmetry \cite{Garcia-Saenz:2018wnw}. 
The massive sector of this multiplet was already studied in \cite{Boulanger:2023lgd} 
using the Becchi-Rouet-Stora-Tyutin-Batalin-Vilkovisky-Stueckelberg   
\cite{Becchi:1975nq,Tyutin:1975qk,Batalin:1981jr,Batalin:1983ggl,BatalinErratum,Stueckelberg:1957zz,Ruegg:2003ps} 
(BRST-BV-Stueckelberg) method developed in \cite{Boulanger:2018dau}. 
The work \cite{Boulanger:2023lgd} led to the discovery of interaction vertices that do not 
vanish in the unitary gauge.
Here we pursue this analysis, adding the massless 
spin-$(1,3/2)$ sector, and investigate whether it is possible to make local 
the rigid (i.e. global) supersymmetry of \cite{Garcia-Saenz:2018wnw} using the 
BRST-BV-Stueckelberg deformation method. 

Since two spin-$3/2$ fields are present in the spectrum, we start with the 
deformations of the gauge algebra corresponding to the $\mathcal{N}=2$ AdS$_4$ 
superalgebra, taking inspiration from the study of $\mathcal{N}=2$ supergravity 
performed in \cite{Boulanger:2018fei}.
We show that there is no non-Abelian deformation of the rigid susy 
algebra of \cite{Garcia-Saenz:2018wnw}.
At best, we are able to exhibit two Abelian couplings between the spin-3/2 sector 
-- where we recall there is one massless and one massive field with adapted mass -- 
and the vector gauge field, see \eqref{InteractingTheoryVertices} below.
These vertices deform the gauge transformations but leave the gauge 
algebra Abelian. They are analogous to couplings present in 
${\cal N}=2$ sugra and exist only for a negative cosmological constant, i.e., 
in AdS$_4$.
We also found a vertex on the de Sitter (dS$_4$) background, that 
couples a pair of real, massive spin-3/2 fields of equal mass with 
a vector field, see \eqref{2_massive_3/2_with_a_vector}. 
These two massive spin-3/2 field form a charged field and the coupling 
to electromagnetism is the minimal one, also present in 
$\mathcal{N}=2\,$ sugra around AdS$_4$, 
but in our case, the cosmological constant can be positive 
at the first order in interactions.  

Then, we add an extra massless gravitino to the spectrum, and 
perform a systematic classification of the possible non-Abelian 
deformations of the free theory that lead to a deformation of the 
Lagrangian. We find that there exists a non-Abelian 
coupling between the PM spin-2 field 
and two massless gravitini in AdS$_4$.
For this part of the work, we use the cohomological method
of \cite{Barnich:1993vg,Henneaux:1997bm}, well-designed for gauge fields.
The non-Abelian vertex we find has recently been obtained by Yu. M. Zinoviev 
\cite{Zinoviev:2024xta} in a different formalism.
We show that the gauge algebra underlying this vertex closely resembles 
the one of $\mathcal{N}=2$ pure supergravity, where the diffeomorphism vector 
is replaced by the gradient of the PM spin-2 gauge parameter.
In fact, with this relation between the translation generator and 
the generator of PM spin-2 transformation, 
we recover all the structures constants of the $\mathcal{N}=2$
supersymmetry algebra on AdS$_4$, 
with the exception of one structure that is identically zero when 
the diffeomorphism vector is a gradient of a scalar. The latter 
term is responsible
for the transformation law of the supercharges under local Lorentz
transformations.
Correspondingly, the non-Abelian cubic vertex that we found is very 
similar to the one of $\mathcal{N}=2$ pure supergravity around AdS$_4$, 
as a comparison with the analysis of \cite{Boulanger:2018fei} shows.

The plan of the paper is as follows. In the next Section \ref{sec:Kinematics}
we spell out our conventions and notation. We also give the free action principles 
for the PM spin-2 field, the massless and massive spin-3/2 fields around 
(A)dS$_4$, as well as the vector gauge field. 
In Section \ref{sec:BRST-Stueck}, we compute the consistent interactions 
among the fields of the spectrum given above, using the BRST-BV-Stueckelberg 
method proposed in \cite{Boulanger:2018dau} and 
further developed in \cite{Boulanger:2023lgd}.
In Section \ref{sec:PM_sugra_vertex}, we use the cohomological 
method of \cite{Barnich:1993vg,Henneaux:1997bm} to classify all the 
possible non-Abelian deformation of the free theory propagating
a PM spin-2 field and two massless gravitini around the AdS$_4$ background.
We present the unique non-Abelian vertex that emerges from the classification
and, using the results of \cite{Boulanger:2018fei}, show the close relation 
between the corresponding gauge-algebra deformation and the gauge algebra 
of ${\cal N}=2$ pure supergravity around AdS$_4$.
In Section \ref{sec:conformalsugra} we show the obstruction of the 
non-Abelian vertex found in the previous section at the level 
of the Jacobi identity, keeping the spectrum of fields unchanged. 
We then add a massless spin-2 field to the spectrum and show that the 
obstruction to the Jacobi identity can be removed thanks to the 
diffeomorphism algebra deformation brought in by the massless spin-2 
field \cite{Boulanger:2000rq}. 
The theory that emerges is ${\cal N}=1$ conformal supergravity, 
here viewed as the only consistent non-Abelian theory that couples 
a partially massless spin-2 field to massless and massive spin-3/2 
fields around AdS$_4$, thereby resolving at the same time the issue 
of the gauging of the rigid supersymmetry carried by the PM multiplet 
of \cite{Garcia-Saenz:2018wnw} and the issue of the consistent 
interactions between a PM spin-2 field and spin-3/2 fields.
Finally, we present our conclusions in Section \ref{sec:conclusions}.

\section{Action principles for free fields in (A)dS}
\label{sec:Kinematics}

In this section, we give our conventions for the (A)dS$_4$ Lorentz-covariant derivatives, and spell out the action principles for the various fields 
under consideration.

In our conventions and notation, $\bar{g}_{\mu \nu}$ denote the metric components 
of the (A)dS$_{4}$ background spacetime with cosmological constant 
\begin{equation}
    \Lambda = -3 \sigma \lambda^2\,, \, \, \, \, \sigma= \pm 1\;,
\end{equation}
where $\lambda$ is the inverse radius of the background 
and the parameter $\sigma=+1$ in the 
AdS$_4$ spacetime, $\sigma=-1$ in the dS$_4$ spacetime. 
On these two backgrounds that we consider in the present article, 
the components of the Riemann tensor read
\begin{equation}
    R_{\mu \nu \rho \sigma} =
    - 2 \sigma \lambda^2 \bar{g}_{\rho[\mu} \bar{g}_{\nu] \sigma}\;,
\end{equation}
where we use the strength-one (anti)symmetrisation convention with 
(square) round brackets around the corresponding indices to be 
(anti)symmetrised. For example,
$F_{\mu\nu} = 2\,\nabla_{[\mu}A_{\nu]}=\nabla_{\mu}A_{\nu}-\nabla_{\nu}A_{\mu}\,$
and $2\,\nabla_{(\mu}\xi_{\nu)} = \nabla_{\mu}\xi_{\nu} + \nabla_{\nu}\xi_{\mu}\,$.

Consequently, on a (co)vector and on a spinor, 
the commutator of (A)dS$_4$ Lorentz-covariant derivatives 
with the Levi-Civita connection is given by
\begin{equation}
\begin{aligned}
    [\nabla_{\mu} , \nabla_{\nu}] V_\rho &= -2\sigma \lambda^2 \,\bar{g}_{\rho[\mu} V_{\nu]}\;,
    \\
    [\nabla_{\mu} , \nabla_{\nu}] \chi_A &= -\frac{\sigma}{2}\,  \lambda^2 (\gamma_{\mu \nu})_A{}^B \chi_B \;,
    \\
\end{aligned}
\end{equation}
where we indicated spinor indices with capital Latin indices. 
In the following, these will be omitted most of the time.
The four Dirac gamma matrices are denoted $\{\gamma_a\}_{a=0,1,2,3}\,$, 
and $\gamma_\mu = \bar{e}_\mu{}^a\,\gamma_a\,$, 
$\gamma_{\mu\nu} = \gamma_{[\mu}\gamma_{\nu]}\,$, with 
$\bar{e}_\mu{}^a$ the components of the (A)dS$_4$ background vierbeins.

\subsection{Partially massless spin-2 field}
\label{PM2kinematics}
We consider a symmetric PM spin-2 field in the Stueckelberg formulation. 
Following the lines of \cite{Boulanger:2018adg}, we allow both signs of the cosmological 
constant, which makes the AdS$_4$ case explicitly non-unitary at the classical level:
\begin{equation}
\begin{aligned}
    L_{\text{PM-St}}  =& -\frac{1}{2} \nabla_\rho h_{\mu \nu} \nabla^\rho h^{\mu \nu }
        +  \nabla_\rho  h^{\mu \nu} \nabla_\mu  h^{\rho}{}_{\nu} 
        -  \nabla_\mu h \nabla_\nu h^{\mu \nu } + \frac{1}{2} \nabla_\mu h \nabla^\mu h + \frac{\sigma}{4} F_{\mu \nu} F^{\mu \nu } \\
        & -2 \sigma \lambda^2 h_{\mu \nu } h^{\mu \nu}  +\frac{1}{2} \sigma \lambda^2 \,h^2
        + 2\lambda~ [h \nabla_\mu B^\mu - h^{\mu \nu } \nabla_\mu B_\nu ] +3 \lambda^2 B^\mu B_\mu\,,
    \label{eq:PM2FreeLagFierzPauliStueck 1}
\end{aligned}
\end{equation}  
where we recall that the parameter $\sigma$ takes the value $+1$ 
in AdS$_4\,$, $-1$ in dS$_4\,$, and where we denoted $h=\bar{g}^{\mu\nu}h_{\mu\nu}\,$.

The above action is invariant under the gauge transformations given by
\begin{equation}
    \left\{ 
        \begin{array}{ll}
           \delta_0 h_{\mu \nu}  & = 2 \,\nabla_{(\mu} \varepsilon_{\nu)} 
           + \lambda \,\bar{g}_{\mu \nu} \,\pi\;, \\[2mm]
           \delta_0 B_{\mu}  & = \nabla_{\mu} \pi + 2 \,\sigma\lambda\, 
           \varepsilon_{\mu}\;.
        \end{array}
    \right.
    \label{eq:PM2GaugeTFS Stueck}
\end{equation}
We can define a tensor that is gauge invariant under the Stueckelberg gauge transformations 
with parameter $\varepsilon_\mu$:
\begin{equation}
    H_{\mu \nu} = h_{\mu \nu} -\frac{\sigma}{\lambda} \nabla_{\!(\mu} B_{\nu)}\;.
\end{equation}
This allows to rewrite the PM spin-2 Lagrangian in the following form, 
where the Stueckelberg field $B_\mu$ appears only through $H_{\mu\nu}$,
\begin{equation}
\begin{aligned}
    L_{\text{PM-St}}  =&  -\frac{1}{2} \nabla_\rho H_{\mu \nu} \nabla^\rho H^{\mu \nu }
        +  \nabla_\rho  H^{\mu \nu} \nabla_\mu  H^{\rho}{}_{\nu} -  \nabla_\mu H \nabla_\nu H^{\mu \nu } + \frac{1}{2} \nabla_\mu H \nabla^\mu H  \\
        & -2 \sigma \lambda^2 H_{\mu \nu } H^{\mu \nu} +\frac{1}{2} \sigma \lambda^2 \, H^2\,.
\label{eq:PM2FreeLagFierzPauliStueck 2}
\end{aligned}
\end{equation}  
If one sets $B_\mu=0$ and $\varepsilon_\mu = -\frac{\sigma}{2 \lambda} \nabla_\mu \pi$ 
in \eqref{eq:PM2FreeLagFierzPauliStueck 2} and then rescale 
$\pi \mapsto -\sigma \lambda \pi$, we recover the usual form for the 
action of the PM spin-2 field \cite{Deser:1983tm, Higuchi:1986py, Deser:2001us}, 
invariant under 
\begin{equation}
    \delta_0 h_{\mu \nu} = \nabla_\mu \nabla_\nu \pi -\sigma \lambda^2\bar{g}_{\mu \nu} \pi \,.
    \label{eq:PM2GaugeTF Unitary gauge}
\end{equation}
It is possible to define an object invariant under the complete gauge transformations 
\eqref{eq:PM2GaugeTFS Stueck}, 
\begin{equation}
    {\cal K}_{\mu \nu \rho}= {\cal K}_{[\mu \nu] \rho} := 2\, \nabla_{[\mu} H_{\nu]\rho}\;,
    \label{eq:CurvPM2Stueck}
\end{equation} 
Using this object, the Lagrangian \eqref{eq:PM2FreeLagFierzPauliStueck 2} 
can be written in the manifestly gauge-invariant way
\begin{equation}
    L_{\text{PM-St}} = -\frac{1}{4} \,{\cal K}_{\mu \nu \rho}\, {\cal K}^{\mu \nu \rho} 
    + \frac{1}{2}\, {\cal K}^\mu \,{\cal K}_\mu \;,
\end{equation}
where ${\cal K}^\mu = \bar{g}_{\nu \rho}{\cal K}^{\mu \nu \rho}\,$.

\subsection{Massless spin-3/2 field}

We describe a Majorana massless spin-3/2 field in AdS$_4$ in terms of a 
spinor-vector $\phi_{\mu}=(\phi_{\mu A})$ and the Rarita-Schwinger 
Lagrangian
\begin{equation}
    L_{3/2} = -\frac{1}{2} \bar{\phi}_\mu \gamma^{\mu\nu\rho}\, 
    \nabla_\nu \phi_\rho 
    + \frac{M}{2}\, \bar{\phi}_\mu \gamma^{\mu\nu} \phi_\nu\;,\quad M=\pm \lambda\;,
    \label{eq:32MasslessLag}
\end{equation}
invariant under the gauge transformations
\begin{equation}
    \delta_0 \phi_\mu = \nabla_\mu \rho + \frac{M}{2}\, \gamma_\mu \,\rho\;,
\end{equation}
where $\rho$ is a Majorana spinor. 
Note the two possible signs in the definition of the mass parameter $M$. 
This is because the gauge invariance condition is quadratic in $M$, 
with $M^2=\lambda^2$. 
It is therefore possible to work with either of the two signs.
This fact will be relevant when coupling a PM spin-$2$ with an even number 
of massless spin-$3/2$ fields. As a matter of a fact, the vertex we found between 
these fields in Section \ref{sec:PM_sugra_vertex}
requires that half the number of spin-$3/2$ have $M=+\lambda$, 
the other half having $M=-\lambda$, 
consistently with the findings of \cite{Zinoviev:2024xta}.
As explained in \cite{Pilch:1984aw}, it is not possible to define a spin-$3/2$ 
field in dS$_4$ spacetime by imposing the Majorana condition 
and at the same time giving it a gauge transformation law of the form 
originally found in \cite{Townsend:1977qa}.
In simpler terms, Majorana massless spin-$3/2$ fields do not 
exist in dS$_4$ spacetime. Defining the gauge-invariant object
\begin{equation}
    \phi_{\mu\nu} = \nabla_{[\mu} \phi_{\nu]} +\frac{M}{2}\, \gamma_{[\mu} \phi_{\nu]}\,,
\end{equation}
the Lagrangian can be rewritten as 
\begin{equation}
    {L}_{3/2} = -\frac{1}{2} \bar{\phi}_\mu \gamma^{\mu \nu \rho}\,  \phi_{\nu\rho} \;, 
    \quad M=\pm \lambda\;.
    \label{eq:32MasslessUSUALGaugeInvLag}
\end{equation}

\subsection{Massive spin-3/2 field}

In the Stueckelberg gauge-invariant formulation, we describe a massive spin-3/2 
field of mass parameter $\omega = \sqrt{\sigma \lambda^2 + m^2}$ 
with a Majorana vector-valued spinor $\psi_{\mu}$ and a Majorana 
spinor Stueckelberg field $\chi$ with Lagrangian 
\begin{equation}
    L_{3/2\text{M}}  =
      - \frac{1}{2} \bar{\psi}_\mu \gamma^{\mu\nu\rho} \nabla_\nu \psi_\rho 
     + \frac{\omega}{2}\,\bar{\psi}_\mu \gamma^{\mu\nu}\psi_\nu
     - \frac{3}{4}  \bar{\chi} \,\gamma^{\mu}\nabla_\mu \chi
     -\frac{3 \omega }{2}\,\bar{\chi}\,\chi    
    -\frac{3 m}{2} \bar{\psi}_\mu \gamma^\mu \chi\;, 
    \label{eq:32MassiveStueckLag}
\end{equation}
which is invariant under the Stueckelberg gauge transformations
\begin{equation}
    \left\{ 
        \begin{array}{ll}
           \delta_0 \psi_{\mu}  & = \nabla_{\mu} \theta 
           + \frac{\omega}{2} \,\gamma_\mu \,\theta\;,  \\[2mm]
           \delta_0 \chi  & = m\, \theta\,.
        \end{array}
    \right.
    \label{eq:32MassiveStueckGaugeTFS}
\end{equation}
Note that, when the mass $m$ parameter is non-zero, 
the massive spin-3/2 field can be defined with the two 
signs of the cosmological constant $\sigma=\pm1$, 
thus in both AdS$_4$ and dS$_4\,$. 
In the dS$_4$ case, the massless limit $m \rightarrow 0$ renders $\omega$ 
imaginary, in clash with the Majorana reality condition. 
This is consistent with the fact that Majorana massless spin-3/2 field cannot be 
defined in dS$_4$, as recalled earlier. 
Defining the gauge invariant field
\begin{equation}
    \Psi_\mu = \psi_\mu - \frac{1}{m}\nabla_\mu \chi 
    - \frac{\omega}{2 m} \gamma_\mu\, \chi\;, 
    \label{eq:32MassiveStueckCurvChi}
\end{equation}
the Lagrangian for the massive spin-3/2 field can be rewritten as 
\begin{equation}
    L_{3/2\text{M}}  =
      - \frac{1}{2} \bar{\Psi}_\mu \gamma^{\mu\nu\rho} \nabla_\nu \Psi_\rho 
     + \frac{\omega}{2}\,\bar{\Psi}_\mu \gamma^{\mu\nu}\Psi_\nu\;. 
    \label{eq:32MassiveStueckLagGaugeInv}
\end{equation}
It is possible to set a gauge, named unitary gauge, where $\chi$ is set to zero using the form of its gauge transformation with parameter $\theta$. 
Once this is done, the gauge invariant field $\Psi_\mu$ coincides with the 
field $\psi_\mu\,$. 

In the unitary gauge, the equations of motion obtained  by taking the Euler-Lagrange 
derivative of the Lagrangian \eqref{eq:32MassiveStueckLagGaugeInv} are 
$\gamma^{\mu \nu \rho} \nabla_\nu \psi_\rho - \omega \gamma^{\mu \nu} \psi_\nu = 0$. 
Using gamma matrices identities and commutators of covariant derivatives of the (A)dS$_4$ 
background, one can verify that this implies:   
\begin{equation}
    (\Box-\omega^2 + 4\sigma \lambda^2) \psi_\mu = 0\;,
    \qquad \omega = \sqrt{\sigma \lambda^2 + m^2}\;.
    \label{Dalembert32}
\end{equation}
In AdS$_4$, when setting the mass parameter $m$ to zero, we recover the gauge-fixed 
equation of motion of a massless spin-3/2 field $(\Box+3\lambda^2)\phi_\mu=0$. 
This equation can be obtained from the massless spin-3/2 Lagrangian \eqref{eq:32MasslessLag} 
in the same way, after a gauge fixing. 
In AdS$_4$ spacetime, note that the parametrisation 
$\omega = \sqrt{\sigma \lambda^2 + m^2}$ does not allow to set $\omega$ to zero 
with a real value of the mass parameter $m$. 
This expresses the fact that the massive spin-3/2 with $\omega=0$ is non-unitary in AdS. 
Indeed, in Anti-de Sitter background, the unitarity bound for the squared mass 
appearing in the D'Alembert-like gauge fixed equation $\Box \phi_\mu = M^2 \phi_\mu$ 
for a Rarita-Schwinger potential is 
$M^2 \geqslant -3\lambda^2\,$. 
Setting $\omega$ to zero gives $\Box\phi_\mu=-4\lambda^2\phi_\mu$, which is below 
the unitarity bound. 

In more general terms, for a spin-$s$ field propagating in 
AdS$_4$, the mass-like term $M^2$ appearing in the 
D'Alembert-like equation $(\Box - M^2 \lambda^2)\phi_s=0\,$ is a 
quadratic function of the conformal dimension $\Delta=E_0$, 
the minimal energy of the corresponding $so(2,3)$ module. 
For a bosonic spin-$s$ field, one has $M^2_{\Delta,s} = \Delta(\Delta-3)-s\,$,
while for a fermionic spin-$s$ field, one has 
$M^2_{\Delta,s} = \Delta(\Delta-3)-s-\frac{1}{4}\,$. 
Massless, propagating spin-$s$ fields in AdS$_4$ have $\Delta = s+1\,$, 
be them bosonic or fermionic 
\cite{Fronsdal:1978rb,Fronsdal:1978vb,Fang:1979hq}. 
These are the propagating Fronsdal and Fang-Fronsdal fields.
The unitarity bound on the conformal dimension 
is $\Delta\geqslant s+1$ for a propagating spin-$s$ field. 
It is saturated by the massless fields.\footnote{The 
spin-$1/2$ and spin-$0$ singletons, called Di and Rac, respectively 
\cite{Flato:1978qz},
are unitary irreducible $so(2,3)$ modules with conformal weights 
$\Delta_{Di}=1\,$ and $\Delta_{Rac}=1/2\,$, respectively, that 
correspond to fields that propagate on the conformal boundary of AdS$_4$.} 
Note that, for $s=0$, the unitary conformally-coupled scalar field with 
$M^2 = -2$ admits not only $\Delta=s+1=1$, but also $\Delta=2$, corresponding 
to two different boundary conditions \cite{Fronsdal:1975eq,Breitenlohner:1982jf}.
The mass-squared $M^2$ is bounded from below by the value taken by the 
massless field.

In this work, we will need a description of the non-unitary massive spin-3/2 field
with $\omega=0$ in AdS$_4$, as it is part of the shortest partially massless 
supermultiplet \cite{Garcia-Saenz:2018wnw} that will be studied in the following. 
We therefore introduce a new parametrisation 
$\Tilde{\omega}=\sqrt{\sigma (\lambda^2- m^2)}$. In AdS$_4$, this parametrisation 
permits to reach only the non-unitary region between $\Box\phi_\mu=-3\lambda^2\phi_\mu$ 
and $\Box\phi_\mu=-4\lambda^2\phi_\mu$. If $m^2>\lambda^2$, $\Tilde{\omega}$ becomes 
imaginary. 
In de Sitter spacetime, $\Tilde{\omega}$ matches with $\omega$, 
and therefore also describe a unitary field, where $m$ can take arbitrary 
values larger than or equal to $\lambda^2$. The Lagrangian consistent with $\Tilde{\omega}=\sqrt{\sigma (\lambda^2-m^2)}$ is
\begin{equation}
    L_{3/2\Tilde{\text{M}}}  =
      - \frac{1}{2} \bar{\psi}_\mu \gamma^{\mu\nu\rho} \nabla_\nu \psi_\rho 
     + \frac{\Tilde{\omega}}{2}\,\bar{\psi}_\mu \gamma^{\mu\nu}\psi_\nu
     + \frac{3\sigma}{4}  \bar{\chi} \,\gamma^{\mu}\nabla_\mu \chi
     + \frac{3 \Tilde{\omega}\sigma }{2}\,\bar{\chi}\,\chi    
     + \frac{3 m\sigma}{2} \bar{\psi}_\mu \gamma^\mu \chi\;.
    \label{eq:32MassiveStueckLagTilde}
\end{equation}
It is unvariant (up to total derivatives) under the 
gauge transformations
\begin{equation}
    \left\{ 
        \begin{array}{ll}
           \delta_0 \psi_{\mu}  & = \nabla_{\mu} \theta 
           + \frac{\Tilde{\omega}}{2} \,\gamma_\mu \,\theta\;,  \\[2mm]
           \delta_0 \chi  & = m\, \theta\,.
        \end{array}
    \right.
    \label{eq:32MassiveStueckGaugeTFSTilde}
\end{equation}
The gauge-invariant Stueckelberg field-strength is
\begin{equation}
    \Psi_\mu = \psi_\mu - \frac{1}{m}\nabla_\mu \chi 
    - \frac{\Tilde{\omega}}{2 m} \gamma_\mu\, \chi\;.
    \label{eq:32MassiveStueckCurvChiTilde}
\end{equation}

\subsection{Massless spin-1 field}

We describe a massless spin-1 field by a vector field $A_\mu$ 
with curvature $G_{\mu\nu}=\nabla_\mu A_\nu - \nabla_\nu A_\mu\,$. 
We recall the expression of the Maxwell Lagrangian,  
\begin{equation}
    L = -\frac{1}{4} G_{\mu\nu} G^{\mu\nu}\;,
\end{equation}
which is invariant under the gauge transformation
\begin{equation}
    \delta A_\mu = \nabla_\mu \alpha \;.
\end{equation}

\section{Cubic deformations using the BRST-BV-Stueckelberg formalism}
\label{sec:BRST-Stueck}

In this section, we briefly summarize the BRST-BV-Stueckelberg method proposed 
in \cite{Boulanger:2018dau}, further developed in \cite{Boulanger:2023lgd}, 
in the context where the set of fields contains a massive spin-3/2 field
and a partially massless spin-2 field, among others. 
More specifically, we consider the fields that build up the shortest 
partially massless supermultiplet of \cite{Garcia-Saenz:2018wnw} and investigate
the possible couplings among those fields.

\subsection{Set-up}
\label{Subsec:SetUpBRSTBVStueckelberg}

We follow the notation of \cite{Garcia-Saenz:2018wnw} as much as possible
and recall the action of the shortest supermultiplet in AdS$_4$ 
with two extra parameters $\sigma$ and $\sigma'$, as compared to what was 
given in \cite{Garcia-Saenz:2018wnw}. 
The first parameter was already introduced above and allows choosing the sign of the cosmological constant, 
while the parameter $\sigma'$ gives us the freedom to change the 
sign of the kinetic terms for the massless spin-3/2 $\phi_\mu$ 
and the vector gauge field $A_\mu$.
The supermultiplet of \cite{Garcia-Saenz:2018wnw} 
is only defined in AdS$_4$ because there is no massless spin-$3/2$ 
fields in dS$_4$ obeying the Majorana reality condition \cite{Pilch:1984aw}, 
as we recalled above. 
Explicitly, the complete free action in the Stueckelberg formulation 
takes the form $S_0  = \int d^4x\sqrt{-\Bar{g}}~ L_0$ with 
\begin{equation}
\begin{aligned}
    L_0  =& ~ -\frac{1}{2} \nabla_\rho h_{\mu \nu} \nabla^\rho h^{\mu \nu }
        +  \nabla_\rho  h^{\mu \nu} \nabla_\mu  h^{\rho}_{~\nu} 
        -  \nabla_\mu h \nabla_\nu h^{\mu \nu } + \frac{1}{2} \nabla_\mu h \nabla^\mu h + \frac{\sigma}{4} F_{\mu \nu} F^{\mu \nu } 
        \\
        & 
        -2 \sigma \lambda^2 h_{\mu \nu } h^{\mu \nu} 
         +\frac{1}{2} \sigma \lambda^2 ~ h^2
        + 2\lambda~ [h \nabla_\mu B^\mu - h^{\mu \nu } \nabla_\mu B_\nu ] 
        +3 \lambda^2 B^\mu B_\mu  \\
        & 
      - \frac{1}{2} \bar{\psi}_\mu \gamma^{\mu \nu \rho} \nabla_\nu \psi_\rho 
     + \frac{\Tilde{\omega}}{2}\bar{\psi}_\mu \gamma^{\mu \nu}\psi_\nu
     + \sigma \frac{3 \Tilde{\omega} }{2} \bar{\chi} \chi +\sigma \frac{3}{4}  \bar{\chi} \gamma^{\mu} \nabla_\mu \chi
     + \sigma \frac{3 m}{2} \bar{\psi}_\mu \gamma^\mu \chi 
     \\
      &+\frac{\sigma'}{2} \bar{\phi}_\mu \gamma^{\mu \nu \rho} \nabla_\nu \phi_\rho - \frac{\sigma'}{2} \lambda  ~\bar{\phi}_\mu \gamma^{\mu \rho} \phi_\rho  +\frac{\sigma'}{4}G^{\mu \nu}G_{\mu \nu}\;.
    \label{FreeTheoryTotalLagrangian}
\end{aligned}
\end{equation}
One recovers the action presented in \cite{Garcia-Saenz:2018wnw}
by taking $\sigma'=1= \sigma$, $\Tilde{\omega}=0$, and setting all the Stueckelberg fields to zero.
We recall the gauge transformations under which the free action is invariant, 
up to boundary terms that we neglect in this work:
\begin{equation}
    \left\{ 
        \begin{array}{ll}
           \delta_0 h_{\mu \nu}  & = 2\,\nabla_{(\mu} \varepsilon_{\nu)} 
           + \lambda \,\bar{g}_{\mu \nu} \,\pi  \\[2mm]
           \delta_0 B_{\mu}  & = \nabla_{\mu} \pi 
           + 2 \,\sigma \lambda\,\varepsilon_{\mu}
        \end{array}
    \right.\quad ,
    \label{GaugeinvHmunuBmu}
\end{equation}
\begin{align}
& \left\{ 
        \begin{array}{ll}
           \delta_0 \psi_{\mu}  & = \nabla_{\mu} \theta + \frac{\Tilde{\omega}}{2}\, 
           \gamma_\mu \,\theta  \\[2mm]
           \delta_0 \chi  & = m \,\theta
        \end{array}
    \right.\quad\;,
    &
    \left\{ 
        \begin{array}{ll}
           \delta_0 \phi_{\mu} &= \nabla_{\mu} \rho + \frac{\lambda}{2}\gamma_\mu \rho  \\[2mm]
           \delta_0 A_{\mu}  &= \nabla_{\mu} \alpha
        \end{array}
    \right.\quad .
\end{align}
For each gauge parameter, we introduce a ghost field with opposite Grassmann parity: 
\begin{itemize}
    \item For $\varepsilon_\nu$, we introduce the Grassmann-odd vector 
    field $\xi_\nu$, and for $\pi$, we introduce the Grassmann odd scalar field $C\,$; 

    \item For $\theta$ we introduce the Grassmann-even Majorana spinor $\zeta\,$; 

    \item For $\rho$ we introduce the Grassmann-even Majorana spinor $\tau\,$; 

    \item For $\alpha$ we introduce the Grassmann-odd scalar field $\epsilon\,$. 
\end{itemize}
The set of all the fields and ghosts will be denoted by $\{\Phi^I\}\,$. 
With each field or ghost $\Phi^I$, one introduces a corresponding 
BV antifield $\widetilde{\Phi}^\star_I$ canonically conjugated to $\Phi^I$ through 
the BV antibracket 
\begin{equation}
\label{BVbracket}
(A, B)=\frac{\delta^R A}{\delta \Phi^I} \frac{\delta^L B}{\delta \widetilde{\Phi}_I^*}
-\frac{\delta^R A}{\delta \widetilde{\Phi}_I^*} \frac{\delta^L B}{\delta \Phi^I} \;,
\end{equation}
for any local functionals $A$ and $B$, and where we use De Witt's condensed 
notations for summations over repeated indices that imply integration over 
spacetime.

The BV functional for the free theory is given by 
\begin{equation}
\begin{aligned}
    W_0 [\Phi, \Phi^{\star}] &=  S_0 
    + \int d^4x ~\sqrt{-\Bar{g}}~ \bigg( h^{\star \mu \nu} ( 2 \,\nabla_{(\mu} \xi_{\nu)} + \lambda \bar{g}_{\mu \nu} C ) 
    + B^{\star \mu} ( \nabla_\mu C + 2 \,\sigma \,\lambda\, \xi_\mu ) \\
    &+ \bar{\psi}^{\star \mu} ( \nabla_\mu \zeta + \frac{\Tilde{\omega}}{2} \gamma_\mu \zeta ) 
    + m\, \bar{\chi}^\star\, \zeta + \bar{\phi}^{\star \mu} ( \nabla_\mu \tau + 
    \tfrac{1}{2}\lambda\, \gamma_\mu \,\tau) + A^{\star \mu} \,\nabla_\mu \epsilon \bigg).
\end{aligned}
\end{equation}
Note that, in our conventions for $W_0\,$, we considered the antifields 
${\Phi}^*_I$ as tensors and not as tensorial densities $\widetilde{\Phi}_I^*$ 
as is understood implicitly in \eqref{BVbracket}. 
Therefore, by defining the BRST differential $s$ for the free theory through 
the BV bracket $s\, \bullet:=\left(W_0, \bullet\right)\,$,  
we have 
$s\,\Phi^I(x) = -\frac{\delta^R W_0}{\delta \widetilde{\Phi}^*_I(x)}
=-\frac{1}{\sqrt{-\bar{g}(x)}}\,\frac{\delta^R W_0}{\delta{\Phi}^*_I(x)}\,$
and 
$s\,\widetilde{\Phi}^*_I(x) = \frac{\delta^R W_0}{\delta {\Phi}^I(x)}=
\sqrt{-\bar{g}(x)}\,(s\,{\Phi}^*_I(x))\,$, hence 
$s\,{\Phi}^*_I(x) = \frac{1}{\sqrt{-\bar{g}(x)}}\,\frac{\delta^R W_0}{\delta{\Phi}^I(x)}\,$.

In the case of a free theory, the BRST differential can be written as the sum 
of two differentials $s = \gamma + \delta\,$, where $\delta$ is the Koszul 
differential that implements the field equations into the BRST cohomology, 
see e.g. the book \cite{Henneaux:1994lbw} for a complete presentation.
The action of these two differentials on the complete BV spectrum is given 
in Table \ref{tableau PM}, where we also list the Grassmann parity, ghost number, 
antifield number and pureghost number of all the fields. 
The problem of perturbative deformation of a free theory, 
in the antifield reformulation of \cite{Barnich:1993vg,Henneaux:1997bm}, 
consists in solving the BV master equation $(W,W)=0$ perturbatively 
around the free solution $W_0$, for the deformed BV-functional 
$W = W_0 + g\,W_1+ {\cal O}(g^2)\,$.
The descent equations issued from the equation $sW_1 = 0$ 
at first order in the formal deformation parameters $g$, 
with $W_1 = \int d^4x \sqrt{-\Bar{g}}~ (a_0 + a_1 + a_2)$
and antifield$(a_i)=i\,$, take the form 
\begin{align}
    \delta a_1 + {\gamma} a_0 &= ~\nabla_\mu j^\mu_0\;, \label{AdSDescentEqa0}\\
    {\delta} a_2 + {\gamma} a_1 &= ~\nabla_\mu j^\mu_1\;, \label{AdSDescentEqa1}\\
    {\gamma} a_2 &= 0\;. \label{AdSDescentEqa2}
\end{align}
Importantly, the cochains $a_0\,$, $a_1\,$, $a_2\,$, 
as well as $j^\mu_0$ and $j^\mu_1$, 
are all assumed to be local. 
All the cohomologies $H^*(s\vert d)$
that we will be computing are in the space of local forms, with representatives 
that depend on the fields and their derivatives up to a finite, but arbitrary, order. 
We refer to Section 2 of \cite{Boulanger:2000rq}
for a detailed presentation of the deformation procedure that we follow.
The term $a_2$ in antifield number two encodes the deformations of the 
Abelian gauge algebra of the free theory, 
the term $a_1$ in antifield number one encodes the 
deformations of the gauge transformations, 
and the antifield number zero part $a_0$ encodes the deformation of the 
quadratic Lagrangian, i.e., the first-order vertices. 
We will make the abuse of terminology of designating by a ``total derivative'' 
a term $\nabla_\mu j^\mu$, it being understood the multiplication with the density 
$\sqrt{-\bar{g}}$ will effectively transform $\nabla_\mu j^\mu$ into a total derivative. 

\begin{table}[t]
\centering
\begin{tabular}{|c||c|c|c|c||c||c|}
\hline
 & $|\cdot|$ & gh & antfld & puregh & $ \gamma$ & $\delta$
\\
\hline 
\footnotesize{$h_{\mu \nu}$} & \footnotesize{$0$} & \footnotesize{$0$} & \footnotesize{$0$} & \footnotesize{$0$} & \footnotesize{$ 2 \nabla_{\!(\mu} \xi_{\nu)} + \lambda\, \bar{g}_{\mu\nu}\, C $} & \footnotesize{$0$} \\ 
\hline 
\footnotesize{$\psi_{\mu }$} & \footnotesize{$1$} & \footnotesize{$0$} & \footnotesize{$0$} & \footnotesize{$0$} & \footnotesize{$- \nabla_{\!\mu} \zeta - \frac{1}{2} 
\tilde{\omega}\,  \gamma_\mu\,\zeta $} & \footnotesize{$0$} 
\\ 
\hline 
\footnotesize{$\phi_{\mu }$} & \footnotesize{$1$} & \footnotesize{$0$} & \footnotesize{$0$} & \footnotesize{$0$} & \footnotesize{$ - \nabla_{\!\mu} \tau - \frac{1}{2} \lambda\, \gamma_\mu\, \tau $} & \footnotesize{$0$} 
\\ 
\hline 
\footnotesize{$A_{\mu }$} & \footnotesize{$0$} & \footnotesize{$0$} & \footnotesize{$0$} & \footnotesize{$0$} & \footnotesize{$ \nabla_{\!\mu} \epsilon $} & \footnotesize{$0$} 
\\ 
\hline \hline
\footnotesize{$B_\mu$} & \footnotesize{$0$} & \footnotesize{$0$} & \footnotesize{$0$} & \footnotesize{$0$} & \footnotesize{$\nabla_{\!\mu} C + 2\sigma \lambda\, \xi_\mu $} & \footnotesize{$0$} \\
\hline 
\footnotesize{$\chi$} & \footnotesize{$1$} & \footnotesize{$0$} & \footnotesize{$0$} & \footnotesize{$0$} & \footnotesize{$ - m ~\zeta$} & \footnotesize{$0$} \\ 
\hline \hline

\footnotesize{$\xi_\mu$} & \footnotesize{$1$} & \footnotesize{$1$} & \footnotesize{$0$} & \footnotesize{$1$} & \footnotesize{$0$} & \footnotesize{$0$} \\
\hline 
\footnotesize{$C$} & \footnotesize{$1$} & \footnotesize{$1$} & \footnotesize{$0$} & \footnotesize{$1$} & \footnotesize{$0$} & \footnotesize{$0$} \\
\hline
\footnotesize{$\zeta$} & \footnotesize{$0$} & \footnotesize{$1$} & \footnotesize{$0$} & \footnotesize{$1$} & \footnotesize{$0$} & \footnotesize{$0$} 
\\ 
\hline 
\footnotesize{$\tau$} & \footnotesize{$0$} & \footnotesize{$1$} & \footnotesize{$0$} & \footnotesize{$1$} & \footnotesize{$0$} & \footnotesize{$0$} \\
\hline
\footnotesize{$\epsilon$} & \footnotesize{$1$} & \footnotesize{$1$} & \footnotesize{$0$} & \footnotesize{$1$} & \footnotesize{$0$} & \footnotesize{$0$} \\
\hline \hline
\footnotesize{$h^{\star \mu \nu}$} & \footnotesize{$1$} & \footnotesize{$-1$} & \footnotesize{$1$} & \footnotesize{$0$} & \footnotesize{$0$}  & \footnotesize{$ \nabla_\rho K^{\rho (\mu \nu )} - \Bar{g}^{\mu \nu} \nabla_{\!\rho} K^{\rho} +\nabla^{(\mu} K^{\nu)} $ } 
\\
\hline 
\footnotesize{$\bar{\psi}^{\star \mu }$} & \footnotesize{$0$} & \footnotesize{$-1$} & \footnotesize{$1$} & \footnotesize{$0$} &  \footnotesize{$0$} & \footnotesize{$ - \nabla_\nu \Bar{\Psi}_\lambda \gamma^{\nu \lambda \mu} - \Tilde{\omega} \Bar{\Psi}_\lambda \gamma^{\mu \lambda}$}
\\
\hline 
\footnotesize{$\bar{\phi}^{\star \mu }$}
& \footnotesize{$0$} &
\footnotesize{$-1$} & \footnotesize{$1$} & \footnotesize{$0$} &  \footnotesize{$0$} & \footnotesize{$\sigma' \bar{\phi}_{\nu \lambda} \gamma^{\mu \nu \lambda}$}
\\
\hline
\footnotesize{$A^{\star \mu}$} & \footnotesize{$1$} & \footnotesize{$-1$} & 
\footnotesize{$1$} & \footnotesize{$0$} & \footnotesize{$0$} & 
\footnotesize{$-\sigma' \nabla_{\!\nu} G^{\nu \mu}$} \\
\hline 
\hline
\footnotesize{$B^{\star\mu}$} & \footnotesize{$1$} & \footnotesize{$-1$} & 
\footnotesize{$1$} & \footnotesize{$0$} & \footnotesize{$0$}  
& \footnotesize{$-2\lambda\, K^\mu$}
\\
\hline
\footnotesize{$\bar{\chi}^{\star }$} & \footnotesize{$0$} & \footnotesize{$-1$} & \footnotesize{$1$} & \footnotesize{$0$} &\footnotesize{$0$} & 
\footnotesize{$\tfrac{3\sigma m }{2 }    \bar{\Psi}_\mu \gamma^\mu$}
\\ 
\hline \hline  
\footnotesize{$\xi^{\star\mu}$} & \footnotesize{$0$} & \footnotesize{$-2$} & $2$ & \footnotesize{$0$} & \footnotesize{$0$} & 
\footnotesize{$-2 \nabla_{\!\alpha} h^{\star \mu\alpha} + 2\sigma \lambda
\,B^{\star\mu}$}
\\
\hline 
\footnotesize{$C^{\star}$} & \footnotesize{$0$} & \footnotesize{$-2$} & $2$ & \footnotesize{$0$} & \footnotesize{$0$} & 
\footnotesize{$\lambda\, h^\star - \nabla_{\!\mu} B^{\star \mu}$}
\\
\hline
\footnotesize{$\bar{\zeta}^{\star}$} & \footnotesize{$1$} & \footnotesize{$-2$} & $2$ & 
\footnotesize{$0$} & \footnotesize{$0$} & 
\footnotesize{$- \nabla_{\!\alpha} \bar{\psi}^{\star \alpha } + \frac{1}{2}\tilde{\omega}\,\bar{\psi}^{\star}_\alpha \, \gamma^\alpha + m\, \bar{\chi}^{\star } $}
\\
\hline
\footnotesize{$\bar{\tau}^{\star}$} & \footnotesize{$1$} &
\footnotesize{$-2$} & $2$ & \footnotesize{$0$} &  \footnotesize{$0$} & \footnotesize{$- \nabla_{\!\mu} \bar{\phi}^{\star \mu} + \frac{1}{2}\lambda\, \bar{\phi}^{\star \mu} \gamma_\mu$} 
\\
\hline 
\footnotesize{$\epsilon^{\star}$} & \footnotesize{$0$} &
\footnotesize{$-2$} & $2$ & \footnotesize{$0$} &  \footnotesize{$0$} & \footnotesize{$- \nabla_{\!\mu} A^{\star \mu}$} \\
\hline
\end{tabular} 
\caption{Properties and BRST differentials of every field and antifield.}
\label{tableau PM}
\end{table}

\subsection{Cohomology of $\gamma$}

The usual procedure is to start computing the cohomology $H^*(\gamma)$ of the 
differential $\gamma\,$, the differential along the gauge orbits.
One finds
\begin{equation}
    H^{*}(\gamma) = \Big\{  f \big(  [\Phi^\star_I ] , [{\cal K}_{\mu \nu \rho}]   , [\Psi_\mu ], [\phi_{\mu \nu}], [G_{\mu \nu}],  C, \nabla_\mu C , \xi_\mu , \nabla_{(\mu} \xi_{\nu)} , \tau , \epsilon  \big) \Big\}\;, \label{eq:CompCohomologyGamma}
\end{equation}
where the notation $[\Phi]$ means the field $\Phi$ and all its derivatives up to some 
finite, but arbitrary, order, so as to ensure the locality of the function $f$.
The quantities in the argument of the function $f$ in the above expression 
are not all independent. First of all, the members of each pair 
$(\nabla_{(\mu} \xi_{\nu)},C)$ and $(\nabla_\mu C,\xi_\mu)$
belong to the same cohomology class, as we discuss below, 
and the differential Bianchi identities
give relations among the derivatives $[{\cal K}_{\mu \nu \rho}]$, 
$[\Psi_\mu ]\,$, $[\phi_{\mu \nu}]\,$, and $[G_{\mu \nu}]\,$.

Details on the calculation of $H^*(\gamma)$ in the sector of 
the massless spin-1 and spin-3/2 fields around AdS$_4$ can be found in 
\cite{Boulanger:2018fei}, for example. 
In the following, we 
detail the contribution of the massive spin-$3/2$ field and of the partially 
massless spin-$2$ field, both in the Stueckelberg formulation. 
It is direct to see that the gauge invariant objects constructed out of the 
fields $h_{\mu\nu},\psi_\mu$ and their Stueckelberg companions 
are ${\cal K}_{\mu \nu \rho}$ and $\Psi_\mu\,$, 
respectively defined  in equations \eqref{eq:CurvPM2Stueck} and 
\eqref{eq:32MassiveStueckCurvChi}. 
Therefore, functions of these gauge invariant objects and their derivatives 
are part of the cohomology of 
$\gamma$: $f([{\cal K}_{\mu \nu \rho}], [\Psi_\mu])   \in H^{*}(\gamma)$.

The ghost fields require further discussion. In the sector of the 
partially massless field $h_{\mu\nu}$, 
there are two ghost fields. One can see in table \ref{tableau PM} 
that both $C$ and $\xi_\mu$ are elements of the cohomology group. 
Indeed, they are 
$\gamma$-closed and there is no way to express them as $\gamma$-exact 
objects. Let us focus on the derivatives of $C$. 
Because of the relation 
$\nabla_\mu C = \gamma B_\mu -2  \sigma \lambda \xi_\mu$, we see that 
$\nabla_\mu C$ is an element of the cohomology in the same class as 
$\xi_\mu$.
At higher orders in derivatives we have the relation 
$\nabla_\alpha \nabla_\mu C = \gamma \nabla_{(\alpha} B_{\mu)} 
-\sigma \lambda \gamma h_{\alpha \mu} + \sigma \lambda^2 \Bar{g}_{\alpha \mu} C$ 
indicating that $\nabla_\alpha \nabla_\mu C$ is in the same cohomology class 
as $C$.
Let us pursue with the derivatives of the ghost $\xi_\mu$. 
When taking the antisymmetric part of the relation 
$\nabla_{\alpha} \nabla_{\mu} C = \gamma \nabla_{\alpha} B_{\mu} - 2 \sigma \lambda \nabla_{\alpha} \xi_{\mu}$, 
one obtains $0 = \gamma \nabla_{[\alpha} B_{\mu]} - 2 \sigma \lambda \nabla_{[\alpha} \xi_{\mu]}$. 
This shows that the antisymmetrized covariant derivative of $\xi_\mu$ is 
$\gamma$-exact, as well as all the derivatives thereof. 
For the symmetric part, the relation 
$\gamma h_{\mu \nu} = 2 \nabla_{(\mu} \xi_{\nu)} + \lambda\, \bar{g}_{\mu\nu} C$ 
indicates that $\nabla_{(\mu} \xi_{\nu)}$ is in the same cohomology class as $C$. 
Taking the covariant derivative of $\gamma h_{\mu \nu}$, one can show that 
$\nabla_\alpha \nabla_{(\mu} \xi_{\nu)}$ is equivalent to $\xi_\mu$ up to 
$\gamma$-exact terms. 
As a conclusion, in the PM spin-2 sector we 
have two cohomology classes given by 
$(\nabla_{(\mu} \xi_{\nu)},C)$ and $(\nabla_\mu C,\xi_\mu)$, 
respectively, as announced above. It will be convenient to take 
one or another representative of each class, depending on the context. 
    
In the sector of the massive spin-3/2 field $\psi_\mu$, 
one can see in Table \ref{tableau PM} that $\zeta$ is $\gamma$-exact, 
and thus it is not an element of the cohomology group, as it should 
for a Stueckelberg field. 
Moreover, because of the relation 
$\nabla_\mu \zeta = - \gamma \psi_\mu - \frac{1}{2} \Tilde{\omega} \gamma_\mu \zeta$, 
$\zeta$ and $\nabla_\mu \zeta$ are related by a $\gamma$-exact term. 
Thus, neither $\zeta$ nor its derivatives belong to the cohomology of $\gamma$. 
This is the typical situation of a massive field theory, as detailed in 
\cite{Boulanger:2018dau}.   
The contributions of the ghost fields to the cohomology of $\gamma$ 
therefore only come from the sectors of the PM spin-2 field $h_{\mu \nu}$, 
of the massless spin-3/2 field $\phi_\mu$, and of the vector gauge field $A_\mu$. 

\subsubsection*{Ambiguity relations}
In a massive theory, the Stueckelberg fields transform under the differential 
$\gamma$ into the ghost fields. We have an example here with the massive spin-$3/2$ 
field whose Stueckelberg field transforms as $\gamma \chi = -m \zeta$. As a 
consequence, the ghost $\zeta$ is not an element of the cohomology group 
and this gives an ambiguity in the 
expression of $\nabla_\mu \zeta$ as a $\gamma$-exact object:
\begin{equation}
    \nabla_\mu \zeta = \gamma(- \frac{1}{m} \nabla_\mu \chi)\;,\quad  
    \nabla_\mu \zeta = \gamma (- \psi_\mu + \frac{\Tilde{\omega}}{2 m} \gamma_\mu \chi)\;,
    \label{Def:Ambizeta}
\end{equation}
as a direct consequence of the gauge invariance of the quantity 
$\Psi_\mu$ defined in \eqref{eq:32MassiveStueckCurvChi}.
Note that, while the second equality of \eqref{Def:Ambizeta} is similar to the case 
of a massless spin-3/2 field, the first equality above exists only 
because of the presence of the Stueckelberg field that transforms into 
the ghost field $\zeta$. 

While we work here in the Stueckelberg formulation, recall that the 
partially massless spin-2 field has a gauge symmetry \eqref{eq:PM2GaugeTF Unitary gauge} 
in the unitary gauge. Consequently, the ghost fields $\xi_\mu\,$, $C$, 
and the derivatives $\nabla_{(\mu}\xi_{\nu)}$ and $\nabla_\mu C$, 
are in the cohomology of $\gamma$. 
Next, we note that one can build an ambiguity relation for 
$\nabla_\rho\nabla_{[\mu}\xi_{\nu]}$ as a $\gamma$-exact quantity,
from the very fact that the quantity ${\cal K}_{\mu\nu\rho}$ defined in 
\eqref{eq:CurvPM2Stueck} is gauge invariant:
\begin{equation}
    \nabla_{\rho} \nabla_{[\mu} \xi_{\nu]} = \gamma(\frac{1}{4\sigma \lambda} \nabla_\rho F_{\mu\nu} ) \;,\quad
    \nabla_{\rho} \nabla_{[\mu} \xi_{\nu]} = \gamma(\nabla_{[\mu} h_{\nu]\rho} + \lambda\, g_{\rho[\mu}B_{\nu]}) \;.
    \label{Def:Ambixi}
\end{equation}
This ambiguity (\ref{Def:Ambixi}) has the same structure as (\ref{Def:Ambizeta}), 
one expression only in terms of the Stueckelberg field and one expression 
in terms of the field and its Stueckelberg partner. 

We make use of these ambiguity relations in the derivation of the cubic 
vertices, so as to ensure that they have a non-singular massless limit  
$m\rightarrow 0$ in the unitary gauge. 
The general strategy in the derivation of vertices 
involving massive fields in the Stueckelerg formulation is detailed in 
\cite{Boulanger:2018dau}. It amounts to building cubic vertices that 
retain the information of the cubic vertices of the massless theory.

\subsection{Cubic deformations of the gauge algebra} \label{Section:choiceOfa2}

In this section, we select the cubic deformations of the gauge algebra 
that are relevant for a localisation of the rigid $\mathcal{N}_4=1$ 
supersymmetry of \cite{Garcia-Saenz:2018wnw}.
First, let us describe a similar theory with which the comparison will be useful. 

Pure $\mathcal{N}_4=2$ supergravity expanded around AdS$_4$ is a theory that, 
in the BRST spectrum, contains a massless graviton with its diffeomorphism 
Grassmann-odd ghost parameter $\xi^\mu$, 
two massless gravitini with Grassmann-even ghost fields $\tau^\Delta$ 
$(\Delta=1,2)$ associated with the Grassmann-odd gauge parameters 
$\rho^\Delta$ for local supersymmetry, and one vector gauge field  
$A_\mu$ with ghost $\epsilon$. 
The gauge transformations on the fields of the spectrum close, on-shell, 
to form the $\mathcal{N}_4=2$ anti-de Sitter superalgebra. This theory has been 
studied using the BRST-BV deformation method in \cite{Boulanger:2018fei}. 
Starting from the free action, it was shown in \cite{Boulanger:2018fei} 
that the cubic deformation of the Abelian algebra leading to $\mathcal{N}_4=2$ 
pure supergravity theory in AdS$_{4}$ is encoded, in the BRST-BV formalism, 
in the following expression at antifield number two,
\begin{equation}
\begin{aligned} 
     a^{\text{sugra}}_2= & \alpha_{3 / 2}\left(\tfrac{1}{4} k_{\Delta \Omega}\xi^{\star \mu}
     \bar{\tau}^{\Delta} \gamma_\mu \tau^{\Omega}+k_{\Delta \Omega} \bar{\tau}^{\star \Delta} \gamma^{\mu \nu} 
     \tau^{\Omega} \nabla_{[\mu} \xi_{\nu]}
     -2 \lambda k_{\Delta \Omega}\bar{\tau}^{\star \Delta} \gamma^\mu \tau^{\Omega} \xi_\mu
     \right) \\ & +y\left(t_{\Delta \Omega}\epsilon^{\star} \bar{\tau}^{[\Delta} 
     \tau^{\Omega]} -2 \lambda t_{\Delta \Omega} \tau^{\star \Delta}
     \tau^{\Omega}\epsilon\right), \quad t_{\Delta \Omega}
     =-t_{\Omega \Delta}, \quad k_{\Sigma \Omega} = k_{\Omega\Sigma }\;.
     \label{Deformation:a2TrainaN2Susy}
\end{aligned}
\end{equation}
By comparing with the PM supermultiplet of \cite{Garcia-Saenz:2018wnw}, 
one can remark that, apart from the fact that the graviton is partially 
massless and that one of the gravitini is massive, there is still one spin-2, 
two spin-3/2 and one spin-1 field. 
This permits to take advantage of the analysis done in 
\cite{Boulanger:2018fei} in the selection of the deformations of the gauge 
algebra we will consider. Indeed, in the PM supermultiplet, there are also two 
fermionic gauge parameters that we shall denote by $\{\rho,\theta\}$: 
the parameter $\rho$ for the massless spin-$3/2$ field $\phi_\mu$, while 
$\theta$ is associated, in the Stueckelberg 
formulation, with the massive spin-$3/2$ field $\psi_\mu\,$. 
The analysis presented in 
the present section consists in the search for 
deformations of the Abelian gauge symmetries, that would correspond 
to a localisation of the $\mathcal{N}_4=1$ rigid supersymmetry of the model. 
Although there are two fermionic gauge parameters, $\theta$ is a Stueckelberg 
gauge parameter that represents an artificial gauge symmetry 
that vanishes in the unitary gauge. 
Therefore, it is expected that, if a localisation of the 
$\mathcal{N}_4=1$ rigid supersymmetry is possible, the gauge parameter 
can only be $\rho$, the gauge parameter of the massless spin-$3/2$ field $\phi_\mu$.  

To make the analysis as complete as possible, we use as starting point the same 
deformations of the gauge algebra as in \eqref{Deformation:a2TrainaN2Susy}
for the analysis of the $\mathcal{N}_4=2$ supergravity. In doing so, the two gauge 
parameters $\rho$ and $\theta$ with corresponding Grassmann-even ghosts 
$\tau$ and $\zeta$ are treated equally, which allows for a clarification 
of their roles in the possible gauging. The differences with 
\eqref{Deformation:a2TrainaN2Susy} is that we do not assume the symmetry properties 
$t_{\Delta \Omega}=-t_{\Omega \Delta}$, $ k_{\Sigma \Omega} = k_{\Omega\Sigma }$ 
provided by the analysis of \cite{Boulanger:2018fei}, and we allow the presence of a $\gamma_5$ matrix as it is present in the global supersymmetry transformations of \cite{Garcia-Saenz:2018wnw}.  
We consider the following list of terms:
\begin{equation}
\begin{aligned}
    a^{\text{Total}}_2 &= k_1\, a_2^{(1)} + k_2\, a_2^{(2)} + k_3\, a_2^{(3)} 
    + k_4\, a_2^{(4)} + k_5\, a_2^{(5)} + k_6\, a_2^{(6)} + k_7\, a_2^{(7)} 
    + k_8\, a_2^{(8)} + k_9\, a_2^{(9)}\\
    & + k_{10}\, a_2^{(10)} + k_{11}\, a_2^{(11)} + k_{12}\, a_2^{(12)} + k_{12'}\, a_2^{(12')} 
    + k_{13}\, a_2^{(13)} + k_{14}\, a_2^{(14)} + k_{15}\, a_2^{(15)}\;,   
\end{aligned}
\end{equation}
with 
\begin{equation}
\begin{aligned}
    a_2^{(1)} &= \xi^{\star \mu} \bar{\zeta} \gamma_\mu \zeta \;,\\
    a_2^{(2)} &= \xi^{\star \mu} \bar{\zeta} \gamma_\mu \tau\; ,\\
    a_2^{(3)} &= \xi^{\star \mu} \bar{\tau} \gamma_\mu \tau\;,\\
    a_2^{(4)} &= \Bar{\zeta}^{\star} \gamma^{\mu \nu} \zeta \nabla_{[\mu} \xi_{\nu]}\;,
\end{aligned}
\quad \quad
\begin{aligned}
    a_2^{(5)} &= \Bar{\zeta}^{\star} \gamma^{\mu \nu} \tau \nabla_{[\mu} \xi_{\nu]}\;,\\
    a_2^{(6)} &= \Bar{\tau}^{\star} \gamma^{\mu \nu} \zeta \nabla_{[\mu} \xi_{\nu]}\;,\\
    a_2^{(7)} &= \Bar{\tau}^{\star} \gamma^{\mu \nu} \tau \nabla_{[\mu} \xi_{\nu]}\;,\\   
    a_2^{(8)} &= \Bar{\zeta}^{\star} \gamma^\mu \zeta \xi_\mu\;,
\end{aligned}
\quad \quad
\begin{aligned}
    a_2^{(9)} &= \Bar{\zeta}^{\star} \gamma^\mu \tau \xi_\mu \;,\\
    a_2^{(10)} &= \Bar{\tau}^{\star} \gamma^\mu \zeta \xi_\mu\;,\\
    a_2^{(11)} &= \Bar{\tau}^{\star} \gamma^\mu \tau \xi_\mu\;,\\
    a_2^{(12)} &= \epsilon^{\star} \bar{\zeta}\tau\;,
\end{aligned}
\quad \quad
\begin{aligned}
    a_2^{(12')} &= \epsilon^\star \bar{\zeta} (i \gamma_5) \tau\;,\\
    a_2^{(13)} &= \Bar{\zeta}^\star \zeta \epsilon\;,\\
    a_2^{(14)} &= \Bar{\zeta}^\star \tau \epsilon\;,\\
    a_2^{(15)} &= \Bar{\tau}^\star \zeta \epsilon\;.
\end{aligned}\label{lista2}
\end{equation}
In this list, the deformations $a_2^{(1)}$, $a_2^{(4)}$, and $a_2^{(8)}$ 
are $\gamma$-exact, hence, cannot deform the gauge algebra.
These were the deformations studied in \cite{Boulanger:2023lgd}, that led 
to two cubic vertices coupling the PM spin-2 field to the massive 
spin-3/2 field. 
Note that there are only four candidates in the above list that are 
in the cohomology of $\gamma\,$: $a_2^{(3)}$, $a_2^{(9)}$, $a_2^{(11)}$, 
and $a_2^{(14)}$. 
All the other ones are $\gamma$-exact terms, and thus can be canceled out by field 
dependent redefinitions of the gauge parameters. They do not deform the gauge algebra. 
Nevertheless, in accordance with the strategy explained in \cite{Boulanger:2018dau}, 
which allows to take the massless limit for the vertices in the unitary gauge, 
the $\gamma$-exact candidates are considered.

\subsection{Deformations of the gauge transformations $a_1$} \label{Sec:DefGaugeTfsa_1}

In this section, we present our results for the solving of the descent 
equation \eqref{AdSDescentEqa1} using the deformations $a_2$ of the gauge algebra 
as sources. Two cases are possible. The first case is when the $a_2$ is part of the 
cohomology of $\gamma$. In that case, either we obtain a solution $a_1$ of the descent 
equation, or either we obtain an obstruction and the deformation computation stops. An 
obstruction is manifest if $\delta a_2$ gives an element of the cohomology of $\gamma$. 
The second case is when $a_2 = \gamma A$. In that case there always exists a trivial 
solution $a_1^t=\delta A$ with $a_0^t=0$. Such a solution is not interesting because the 
theory is not really deformed, $a_2$ and $a_1^t$ can always be canceled out by field 
dependent redefinitions of the gauge parameters and quadratic 
redefinitions of the fields in the quadratic Lagrangian. 

However, since we work with massive and partially massless fields in the Stueckelberg 
formulation, it is in general possible to obtain an alternative solution $a_1$ that differs 
from $a_1^t$ by a $\gamma$-exact term $\gamma c_1$ plus a term $\bar{a}_1$ 
in the cohomology of $\gamma\,$:
\begin{equation}
    a_1 - a_1^t = \gamma c_1 + \bar{a}_1\;,\qquad \bar{a}_1\in H(\gamma)\;.
\end{equation}
These non-trivial solutions $a_1$ are obtained by making use of the 
right-hand sides of the ambiguity relations \eqref{Def:Ambixi} and 
\eqref{Def:Ambizeta}, as well 
as a certain number of integrations by parts. 
The term $\bar{a}_1$ is a non-trivial deformation 
of the gauge transformations, that cannot be removed by field-dependent 
redefinition of the gauge parameter, and that could lead to a non-trivial 
interaction vertex.

At this stage, we obtain that all the non-trivial deformations 
of the gauge algebra, $a_2^{(3)}$, $a_2^{(9)}$, $a_2^{(11)}$ and $a_2^{(14)}$, 
are obstructed. Apart from $a_2^{(3)}$, $a_2^{(9)}$, $a_2^{(11)}$ and $a_2^{(14)}$, 
all the other candidates that survive are $\gamma$-trivial, hence do 
not deform the gauge algebra.
This means that, if we find vertices at the next stage, they will be Abelian. 
Therefore, we can already conclude that there is 
no way to localise the rigid supersymmetry carried by the shortest 
PM supermultiplet of \cite{Garcia-Saenz:2018wnw}, without adding any extra 
fields to the spectrum.

As an example, starting from the $\gamma$-exact deformation  
$a_2^{(12)}=\epsilon^{\star} \,\bar{\zeta}\,\tau= \gamma \left( -\frac{1}{m} \epsilon^\star\, \bar{\chi} \,\tau \right)\,$, 
we obtain the results
        \begin{equation}
            a_1^{(12)} = A^{\star \mu} \Bigl( - \bar{\psi}_\mu \tau + \bar{\phi}_\mu \zeta - \Bigl(\frac{\Tilde{\omega}-\lambda}{2 m}\Bigr)\bar{\chi}\gamma_\mu \tau \Bigr),
        \end{equation}
        \begin{equation}
        \begin{aligned}
            c_1^{(12)} &=  -\frac{1}{m} A^{\star \mu}\,  \bar{\chi}\, \phi_\mu, ~~~\bar{a}_1^{(12)} =  A^{\star \mu} \,\bar{\tau}\,\Psi_\mu\;.
        \end{aligned}
        \end{equation}

\subsection{Non-trivial cubic interaction vertices $a_0$}
In this section, we pursue the deformation procedure with the $a_1$ terms 
that appeared at the previous stage, coming form the surviving $a_2$ candidates. 
These $a_1$'s are now used as sources for the $a_0$ in Equation 
\eqref{AdSDescentEqa0}. At this level, we can add deformations of the gauge transformations $\bar{a}^{(I)}_1$ 
solutions of $\gamma \bar{a}^{(I)}_1 =0$. Only those $\bar{a}^{(I)}_1$ that are nontrivial
representatives of $H(\gamma)$ will be considered. 
Such deformations solve \eqref{AdSDescentEqa1} with an $a_2=0$.

For each candidate $a_1$ obtained at the previous section, with $\bar{a}_1\neq0$, 
the goal is to find a combination $\alpha^{(A)} \bar{a}^{(A)}_1 + \alpha^{(B)} \bar{a}^{(B)}_1 + \alpha^{(C)} \bar{a}^{(C)}_1 +\dots$ of elements $\bar{a}^{(I)}_1$ in the cohomology of $\gamma$ with coefficients $\alpha^{(I)}$ such that
\begin{equation}
\begin{aligned}
    \delta (\bar{a}_1^{Tot}) +\gamma a_0 &=  \text{t.d.}\, ,\\
    \bar{a}_1^{Tot} &= \bar{a}_1 + \alpha^{(A)} \bar{a}^{(A)}_1 + \alpha^{(B)} \bar{a}^{(B)}_1 + \alpha^{(C)} \bar{a}^{(C)}_1 +\dots \, .
\end{aligned}  
\end{equation}
Then, to obtain the final result, we define ${a}_1^{Tot}$, 
which is just $a_1$ but with $\bar{a}_1$ substituted with $\bar{a}_1^{Tot}$, 
\begin{equation}
    a_1^{Tot} = a_1^t + \gamma c_1 + \bar{a}_1^{Tot} \, .
\end{equation}
Injecting $a_1^{Tot}$ in the descent equation \eqref{AdSDescentEqa0}, we obtain the complete vertex
\begin{equation}
    a_0^{Tot} = a_0 + \delta c_1 \, .
\end{equation}
This solution $a_0^{Tot}$ is a non-trivial deformation provided that 
$a_0 \neq \delta b$.  
Indeed, if $a_0 = \delta b$ for some local function $b$, 
this means that $\bar{a}_1^{Tot}= \gamma c$. Although 
$\bar{a}_1^{Tot}$ is a sum of elements, each of which non-trivial in the cohomology 
of $\gamma$, it may happen that the sum is trivial in the cohomology, because of 
some cancellations. 
Therefore, in that case where $a_0 = \delta b$, the deformation is trivial 
and can be removed by field redefinitions.

In the following, we present our results for the only candidates $a_1$ 
that lead to interaction vertices $a_0$. We found three vertices 
denoted $a_0^{(12)}$, $a_0^{(12')}$ and $a_0^{(30)}$. 

\subsubsection{Interaction vertex $a_0^{(12)}$}
Starting with the trivial deformation of the gauge algebra 
$a_2^{(12)}= \epsilon^\star\, \bar{\zeta}\, \tau$, we find a 
deformation $a_1^{Tot(12)}$ of the gauge transformations 
that gives rise to a vertex $a_0^{(12)}$
provided that $\Tilde{\omega}=\sqrt{\sigma(\lambda^2-m^2)}$ vanishes. 
Explicitly, we have  
\begin{equation}
\begin{aligned}
    a_1^{Tot(12)} =\;& A^{\star \mu} \Bigl( - \bar{\psi}_\mu\, \tau 
    + \bar{\phi}_\mu \,\zeta 
    + \tfrac{\lambda}{2 m}\,\bar{\chi}\,\gamma_\mu \,\tau \Bigr) \\
    &+ \bar{\psi}^{\star\mu} \Bigl( \tfrac{\sigma'}{4}\, \gamma_{\mu \rho \sigma}\, \tau\; 
    G^{\rho \sigma} - \tfrac{\sigma'}{2}\, \gamma^\beta \,\tau \,G_{\mu \beta} 
    - \tfrac{\sigma'}{6 \lambda}\, \nabla_\mu\! \left(\gamma^{\rho \sigma}\, 
    \tau \,G_{\rho \sigma}\right)\Bigr)\, ,
\end{aligned}
\end{equation}
and 
\begin{equation}
    a_0^{(12)} = \sigma' \bar{\Psi}_\beta \Bigl( G^{\alpha \beta} - i \gamma^5 (*G)^{\alpha \beta}\Bigr) \phi_\alpha
    + \frac{\sigma'}{m} \nabla_{\!\nu} G^{\nu \mu}\, \bar{\chi}\, \phi_\mu\;,
    \label{a012}
\end{equation}
where we have used that 
$\gamma^{\mu  \nu \rho \sigma} = 
-i \gamma^5 \epsilon^{\mu \nu \rho \sigma}$ and $(*G)^{\mu \nu} 
= \frac{1}{2} \epsilon^{\mu \nu \rho \sigma} G_{\rho \sigma}$.
This is a new nontrivial vertex involving a massless spin-3/2 field, 
a massive spin-3/2 field of mass parameter $\Tilde{\omega}=0$ and a massless spin-1 field. 
The gauge transformations are deformed, but the gauge algebra is not. 
This reproduces a vertex of the $\mathcal{N}_4=2$ supergravity theory in AdS$_4$, except 
that here one of the two spin-3/2 fields is massive -- see Equation (3.48) of 
\cite{Boulanger:2018fei}.

\subsubsection{Interaction vertex $a_0^{(12')}$}

Starting with the candidate 
$a_2^{(12')} = \epsilon^\star \bar{\zeta} (i \gamma_5) \tau\,$, 
we obtain a consistent vertex, provided that the mass parameter 
$\Tilde{\omega}=\sqrt{\sigma(\lambda^2-m^2)}$ vanishes. 
Explicitly, 
\begin{equation}
\begin{aligned}
    a_1^{Tot(12')} =\;& A^{\star \mu} \Bigl( - \bar{\psi}_\mu (i \gamma_5 )\tau + \bar{\phi}_\mu (i \gamma_5 )\zeta - \frac{\lambda}{2 m}\bar{\chi}\gamma_\mu(i \gamma_5 ) \tau \Bigr) \\
    & -\frac{\sigma'}{2} \Bigl\{\bar{\psi}^{\star\rho} \Bigl( \gamma^\beta (i \gamma_5) \tau G_{\rho \beta}-\frac{1}{2} \gamma_\rho{}^{\alpha \beta} (i \gamma_5) \tau G_{\alpha \beta}\Bigr) \\
& +\bar{\chi}^\star \Bigl(\tfrac{\sigma\lambda}{3 m} \gamma^{\alpha \beta} (i \gamma_5) \tau G_{\alpha \beta}\Bigr)\Bigr\}\;,
\end{aligned}
\end{equation}
and 
\begin{equation}
    a_0^{Tot(12')} = \sigma' \bar{\Psi}_\sigma (i \gamma_5)\Bigl[  G^{\rho \sigma} -  i \gamma_5(* G)^{\rho \sigma}\Bigr] \phi_\rho + \frac{\sigma'}{m} \nabla_\nu G^{\nu \mu} \bar{\chi} (i \gamma_5) \phi_\mu
\end{equation}
This is a new non-trivial vertex involving a massless spin-3/2 field, 
a massive spin-3/2 field with mass parameter $\Tilde{\omega}=0$ and a massless spin-1 field. 
The gauge transformations are deformed, but the gauge algebra is not. The vertex is very 
similar to the vertex $a_0^{(12)}$ above in \eqref{a012}.

\subsubsection{Interaction vertex $a_0^{(30)}$}

In this section, we allow several massive spin-3/2 fields to couple to a gauge 
vector. Such states can generically appear in the supermultiplets of 
\cite{Bittermann:2020xkl,Bobev:2021oku}, for example.
We therefore consider the free Lagrangian for $N>1$ massive,  
Majorana spin-$3/2$ fields $\psi_\mu^\Omega$ $(\Omega=1, \dots, N)$ 
with diagonal mass matrix $\omega_{\Delta\Omega}$,  
added to the Lagrangian for a gauge field $A_\mu$ with field strength $G_{\mu\nu}\,$:
\begin{equation}
    L_0  =- \frac{1}{2} \bar{\Psi}_\mu^{\Delta} \gamma^{\mu \nu \rho} \nabla_\nu \Psi_\rho^{\Omega} \delta_{\Delta \Omega} 
     + \frac{1}{2}\omega_{\Delta\Omega }\bar{\Psi}_\mu^{\Delta} \gamma^{\mu \nu}\Psi_\nu^{\Omega}
     +\frac{\sigma'}{4}G^{\mu \nu} G_{\mu \nu}\;.
\end{equation}
We will not need to impose the mass parameters on the diagonal of $\omega_{\Delta\Omega}$ 
to vanish, therefore we use the parametrisation that describes the unitary region 
in AdS$_4$ spacetime: $\omega_{\Delta \Delta} =\sqrt{\sigma \lambda^2 + m_{\Delta \Delta}^2}\,$, where the index $\Delta$ is fixed.
We consider the following trivial deformation of the gauge algebra, 
which is a generalization of $a_2^{(13)}$ in \eqref{lista2}: 
\begin{equation}\label{a230}
    a_2^{(30)}= \bar{\zeta}^\star_\Delta \,\zeta_\Omega\, \epsilon\, t^{\Delta \Omega}\;.
\end{equation}
This deformation $a_2^{(30)}$ can be lifted to a deformation of the gauge 
transformations, that in turn can be integrated to a consistent vertex provided 
$\omega_{\Delta \Omega}=\omega\delta_{\Delta \Omega}$ and $t^{\Delta \Omega}
=t^{[\Delta \Omega]}\,$. 
The resulting deformation of the gauge transformations and the corresponding 
cubic vertex read
\begin{equation}
 a_1^{(30)}=\bar{\psi}_{\Delta}^{\star\alpha}\, (\psi_{\Omega \alpha}\, \epsilon\, 
 t^{\Delta \Omega}- \zeta_{\Omega}\, A_\alpha\, t^{\Delta \Omega} )
 + \bar{\chi}_{\Delta}^\star\, \chi_{\Omega}\, \epsilon \,t^{\Delta \Omega} \; , 
\end{equation}
\begin{equation}\label{vertexa30}
    a_0^{(30)} = -\frac{1}{m} A_\alpha \bar{\Psi}_{\Delta \mu \nu} \gamma^{\alpha \mu \nu} \chi_\Omega \,t^{\Delta \Omega} - \frac{1}{2}  A_\nu\bar{\Psi}_{\Delta\lambda} \gamma^{\alpha \nu \lambda} \Psi_{\Omega \alpha} t^{\Delta \Omega}\; .
\end{equation}
where we recall that 
$\Psi^\Delta_\mu = \psi^\Delta_\mu - \frac{1}{m}\nabla_\mu \chi^\Delta 
    - \frac{\omega}{2 m} \gamma_\mu\, \chi^\Delta\,$, 
    and $\Psi_{\mu \nu}^\Delta=\nabla_{[\mu}\Psi^\Delta_{\nu]}+\frac{\omega}{2}\gamma_{[\mu} \Psi^\Delta_{\nu]}\,$.

This is, in the Stueckelberg formalism, the well-known minimal coupling of 
massive spin-3/2 fields to electromagnetism. The simplest case is $N=2\,$, 
where the pair of Majorana Rarita-Schwinger field forms a $U(1)$ doublet, 
a complex Rarita-Schwinger field, and where one can take the matrix 
$t^{\Delta \Omega}=\epsilon^{\Delta \Omega}\,$, the $so(2)$-invariant 
symbol. 

This problem has an old story, of course \cite{Velo:1969bt, Johnson:1960vt}. 
Let us only mention the analyses \cite{Deser:2000dz,Deser:2001dt} that went 
beyond formal consistency, looking at causality of the models, in particular. 
There, general Pauli-type non-minimal couplings of the 
type \eqref{a012} are added, on top of the minimal couplings, to ensure the correct 
number of propagating degrees of freedom. It is of course not a surprise that, 
among the 5-parameter family of non-minimal vertices considered in 
\cite{Deser:2000dz,Deser:2001dt}, precisely the ones corresponding to 
\eqref{a012} (but for equal-mass spin-3/2 fields) emerge on the account of counting 
of degrees of freedom. In the Stueckelberg formalism, we see that these non-minimal 
couplings appear automatically. It is one of the advantages of the Stueckelberg 
formalism, that it naturally account for the correct propagation of degrees 
of freedom. Note that in flat space, this coupling was also studied in 
\cite{Zinoviev:2006im}. 

\subsection{Final results}
In this section, we present our final results outside 
the BRST-BV formalism and in the unitary gauge, 
where all the Stueckelberg fields have been eliminated. The procedure to reach the unitary gauge at first order in deformation is detailed in \cite{Boulanger:2023lgd}.

\subsubsection{Vertices $a_0^{(12)}$ and $a_0^{(12')}$}
Our two results, applying to the free supermultiplet of \cite{Garcia-Saenz:2018wnw}, and 
that exist only when $\Tilde{\omega}=0$ are the following. Consider a model composed of one massive 
spin-3/2 field with mass parameter $\Tilde{\omega}=0$, one massless spin-3/2 field and one massless 
spin-1 field, in AdS$_4$ spacetime. The Lagrangian takes the form
\begin{equation}
\begin{aligned}
    L_0  =
      - \frac{1}{2} \bar{\psi}_\mu \gamma^{\mu \nu \rho} \nabla_{\!\nu} \psi_\rho 
      +\frac{\sigma'}{2} \bar{\phi}_\mu \gamma^{\mu \nu \rho} \nabla_{\!\nu} \phi_\rho 
      - \frac{\sigma'}{2} \lambda  ~\bar{\phi}_\mu \gamma^{\mu \rho} \phi_\rho  +\frac{\sigma'}{4}G^{\mu \nu}G_{\mu \nu}\;.
    \label{InteractingTheoryFreeLag}
\end{aligned}   
\end{equation}
The action is invariant under the gauge transformations
\begin{equation}
     \delta_0 \phi_{\mu} = \nabla_{\mu} \rho + \frac{\lambda}{2} \gamma_\mu \rho\,, \;\; \delta_0 A_{\mu}  = \nabla_{\mu} \alpha\,, \;\; \delta_0 \psi_\mu = 0\,.
\end{equation}
Our result is that the deformed action with Lagrangian $L_0 + L_1$, 
\begin{equation}
\begin{aligned}
    L_1  = g^{(12)} \sigma' \bar{\psi}_\beta \Bigl[ G^{\alpha \beta} - i \gamma_5 (*G)^{\alpha \beta}\Bigr] \phi_\alpha + g^{(12')}\sigma' \bar{\psi}_\sigma (i \gamma_5)\Bigl[  G^{\rho \sigma} -  i \gamma_5(* G)^{\rho \sigma}\Bigr] \phi_\rho\,,
    \label{InteractingTheoryVertices}
\end{aligned}   
\end{equation}
is invariant under the deformed gauge transformations $\delta_0 + \delta_1$, with  
\begin{equation}
     \delta_1 A_\mu = - g^{(12)} \bar{\psi}_\mu \rho - g^{(12')} \bar{\psi}_\mu (i \gamma_5) \rho\,,
\end{equation}
\begin{equation}
\begin{aligned}
    \delta_1 \psi_\mu =~& \sigma' g^{(12)}\Bigl[ \tfrac{1}{4} \gamma_{\mu \nu\rho}\, 
    \rho \,G^{\nu\rho} - \tfrac{1}{2} \gamma^\nu \rho\, G_{\mu \nu} 
    - \tfrac{1}{6 \lambda} \nabla_{\!\mu} \Bigl( \gamma^{\rho \sigma} \rho\, G_{\rho \sigma} \Bigr) \Bigr] \\ 
    +~&\sigma' g^{(12')}\Bigl[ \tfrac{1}{4} \gamma_\mu{}^{\alpha \beta}(i \gamma_5) \rho\, G_{\alpha \beta}  - \tfrac{1}{2} \gamma^\nu (i \gamma_5) \rho\, G_{\mu \nu} + \tfrac{\sigma}{6 \lambda} \nabla_{\!\mu} \Bigl(\gamma^{\alpha \beta} (i \gamma_5)
    \,\rho\, G_{\alpha \beta}\Bigr) \Bigr]\,,
\end{aligned}
\end{equation}
\begin{equation}
    \delta_1 \phi_\mu =0\,.
\end{equation}
The gauge algebra remains Abelian.

\subsubsection{Vertex $a_0^{(30)}$}
Consider a model composed of an even number $N$ of massive spin-3/2 fields $\psi_\mu^\Omega$, $\Omega=1,\dots,N$ of same mass $\omega$ and a massless spin-1 field $A_\mu$. This model is defined in (A)dS$_4$.
The free Lagrangian takes the form 
\begin{equation}
    L_0  =- \frac{1}{2} \bar{\psi}_\mu^{\Omega} \gamma^{\mu \nu \rho} \nabla_\nu \psi_\rho^{\Delta} \delta_{\Omega \Delta} 
     + \frac{\omega}{2}\bar{\psi}_\mu^{\Omega} \gamma^{\mu \nu}\psi_\nu^{\Delta} \delta_{\Omega \Delta}
     +\frac{\sigma'}{4}G^{ \mu \nu} G_{\mu \nu}\;.
\end{equation}
The action is invariant under the free gauge transformations 
\begin{equation}
    \delta_0 A_\mu = \nabla_\mu \alpha\,.
\end{equation}
Our result is that the action with deformed Lagrangian $L_0+ L_1$,
\begin{equation}
\label{2_massive_3/2_with_a_vector}
    L_1 = -g^{(30)} \frac{1}{2} A_\nu \bar{\psi}_{\Delta \lambda} \gamma^{\alpha \nu \lambda} \psi_{\Omega \alpha} t^{\Delta \Omega}\,,\; \; t^{\Delta \Omega} =  t^{[\Delta \Omega]} \,,
\end{equation}
is invariant under the deformed gauge transformations $\delta_0+\delta_1$ with 
\begin{equation}
    \delta_1 \psi^\Delta_\mu= g^{(30)} \psi_{\Omega \mu} \alpha t^{\Delta \Omega}\;,\qquad \delta_1 A_\mu = 0\;.
\end{equation}
The gauge algebra remains Abelian.
We recover of course the well known minimal coupling between a doublet of 
Majorana massive spin-3/2 fields and a massless spin-1 field 
\cite{Deser:2000dz, Deser:2001dt, Zinoviev:2006im}.

\section{Partially massless supergravity vertex}
\label{sec:PM_sugra_vertex}

In this section, we push further the analogy with ${\cal N}=2$
pure supergravity around AdS$_4$ by adding at least one extra massless gravitino 
to the spectrum. 
We will classify all the possible deformations of the gauge algebra 
and obtain a non-Abelian vertex that bears striking resemblance 
with the gravitational minimal coupling in ${\cal N}=2$ sugra, 
except that the role of the graviton is taken over by the 
PM spin-2 field. 

\subsection{Set-up}
We consider a free theory consisting of a PM spin-2 field $h_{\mu \nu}$ 
in the unitary gauge and an unspecified number of massless spin-3/2 fields 
$\phi_{\mu}^\Delta$, $\Delta=1,\dots,N$, with $N>1\,$ so that at least 
two massless gravitini are present, which ensures that one can extract 
the spectrum of ${\cal N}=2$ pure supergravity from the total spectrum, 
recalling that a partially massless spin-2 field decomposes into a 
massless spin-2 and a massless spin-1 fields in the flat spacetime limit.
The total, free Lagrangian, takes the form
\begin{equation}
    L_0 = -\frac{1}{4} K_{\mu \nu \rho} K^{\mu \nu \rho} + \frac{1}{2} K^\mu K_\mu -\frac{1}{2} \bar{\phi}_\mu^{\Delta} \gamma^{\mu \nu \rho} \nabla_\nu \phi_\rho^{\Sigma} \delta_{\Delta \Sigma} + \frac{\lambda}{2} \bar{\phi}_\mu^{\Delta} \gamma^{\mu \rho} \phi_\rho^{\Sigma} M_{\Delta \Sigma}\,,
    \label{PMSugraFinalResL0}
\end{equation}
where $K_{\mu\nu\rho}=2\,\nabla_{\![\mu}h_{\nu]\rho}$ and 
$K_\mu = \bar{g}^{\nu\rho}K_{\mu\nu\rho}\,$, with gauge transformations
\begin{align}
\delta_0 h_{\mu \nu} &=   \nabla_\mu \nabla_\nu \pi -\sigma \lambda^2\bar{g}_{\mu \nu} \pi  \;,\\
\delta_0 \phi_{\mu}^\Delta &=    \nabla_\mu \rho^\Delta +\frac{\lambda}{2} \gamma_\mu \rho^\Sigma M_\Sigma{}^{\Delta}  \;.
\end{align}
The mass matrix $M_{\Delta \Sigma}=\text{diag}(\pm 1,\pm 1, \dots, \pm 1)$, 
$M_{\Delta \Gamma}M^{\Gamma}{}_{\Sigma}=\delta_{\Delta \Sigma}$, allows for massless 
spin-$3/2$ fields with opposite mass-like terms. We note that, in order to have an 
invariance under the R-symmetry group $O(N)$, one needs $M_{\Delta \Sigma}$ 
to correspond to an invariant tensor of $O(N)$. 
The only possibility is $M_{\Delta \Sigma}=\pm \delta_{\Delta \Sigma}$. 
Therefore, in the case $M_{\Delta \Sigma}\neq\pm \delta_{\Delta \Sigma}$, the R-symmetry 
$O(N)$ is explicitly broken from the start. Note also that for $N>2$, say $N=4$, 
we can construct an $M_{\Delta \Sigma}\neq\pm \delta_{\Delta \Sigma}$ 
breaking $O(4)$, but that is 
a direct sum of invariant tensors of $O(2)$: 
$M_{\Delta \Sigma}=\pm\text{diag}(- 1,- 1, +1,+ 1)$.

As in the previous section, we introduce a ghost field $\tau^\Delta$ associated to each gauge parameter $\rho^\Delta$ and a ghost field $C$ associated with the PM gauge parameter $\pi$. The set of all fields and ghost fields is denoted ${\Phi^I}$. To each field $\Phi^I$ is associated an antifield $\Tilde\Phi^{\star}_I$ canonically conjugated to $\Phi^I$ by the BV antibraket defined at \eqref{BVbracket}. The complete list of fields and antifields as well as their BRST differentials are summarized in table \ref{tableau PM unitary gauge}. The BV functional of the free theory takes the form 
\begin{equation}
\begin{aligned}
    W_0 [\Phi, \Phi^{\star}] &=  S_0 
    + \int d^4x \,\sqrt{-\Bar{g}}\, \big( h^{\star \mu \nu} ( \nabla_\mu \nabla_\nu C - \sigma \lambda^2 \bar{g}_{\mu \nu} C ) 
    + \bar{\phi}^{\star \mu}_\Delta ( \nabla_\mu \tau^\Delta + 
    \tfrac{\lambda}{2} \gamma_\mu \tau^\Sigma M_\Sigma{}^\Delta) \big).
\end{aligned}
\end{equation}
The cohomology group of the differential along the gauge orbits $\gamma$ is 
\begin{equation}
    H^{*}(\gamma) = \Big\{  f \big(  [\Phi^\star_I ] , [K_{\mu \nu \rho}] , [\phi^\Delta_{\mu \nu}],  C, \nabla_\mu C, \tau^\Delta \big) \Big\}\;. \label{eq:CompCohomologyGammaPMSUGRA}
\end{equation}
As already discussed in subsection \ref{Subsec:SetUpBRSTBVStueckelberg}, we search for perturbative deformations of the BV functional at first order, $W=W_0+ g W_1+\mathcal{O}(g^2)$, $W_1 = \int d^4x \sqrt{-\Bar{g}}~ (a_0 + a_1 + a_2)$, by solving classical master equation $(W,W)=0$ to first order that take form of the descent of equations
\begin{align}
    \delta a_1 + {\gamma} a_0 &= ~\nabla_\mu j^\mu_0\;, \label{AdSDescentEqa0PMSUGRA}\\
    {\delta} a_2 + {\gamma} a_1 &= ~\nabla_\mu j^\mu_1\;, \label{AdSDescentEqa1PMSUGRA}\\
    {\gamma} a_2 &= 0\;. \label{AdSDescentEqa2PMSUGRA}
\end{align}
We recall that the terms of antifield number 2, $a_2$, are deformations of the gauge algebra, terms of antifield number 1, $a_1$, deformations of the gauge transformations, and terms of antifield number 0, $a_0$, deformations of the free action.
\begin{table}[t]
\centering
\begin{tabular}{|c||c|c|c|c||c||c|}
\hline
 & $|.|$ & gh & afld & puregh & $\gamma$ & $\delta$
\\
\hline 
\footnotesize{$h_{\mu \nu}$} & \footnotesize{$0$} & \footnotesize{$0$} & \footnotesize{$0$} & \footnotesize{$0$} & \footnotesize{$\nabla_\mu \nabla_\nu C - \sigma \lambda^2 \bar{g}_{\mu \nu} C $} & \footnotesize{$0$} \\ 
\hline 

\footnotesize{$\phi_{\mu}^\Delta$} & \footnotesize{$1$} & \footnotesize{$0$} & \footnotesize{$0$} & \footnotesize{$0$} & \footnotesize{$ - \nabla_{\mu} \tau^\Delta - \frac{\lambda}{2}  \gamma_\mu \tau^\Omega M_\Omega{}^\Delta $} & \footnotesize{$0$} 
\\ 
\hline \hline
 
\footnotesize{$C$} & \footnotesize{$1$} & \footnotesize{$1$} & \footnotesize{$0$} & \footnotesize{$1$} & \footnotesize{$0$} & \footnotesize{$0$} \\
\hline
\footnotesize{$\tau^\Delta$} & \footnotesize{$0$} & \footnotesize{$1$} & \footnotesize{$0$} & \footnotesize{$1$} & \footnotesize{$0$} & \footnotesize{$0$} \\
\hline
\hline
\footnotesize{$h^{\star \mu \nu}$} & \footnotesize{$1$} & \footnotesize{$-1$} & \footnotesize{$1$} & \footnotesize{$0$} & \footnotesize{$0$}  & \footnotesize{$ \nabla_\rho K^{\rho (\mu \nu )} - \Bar{g}^{\mu \nu} \nabla_\rho K^{\rho} +\nabla^{(\mu} K^{\nu)} $ } 
\\
\hline 
\footnotesize{$\bar{\phi}^{\star \mu }_\Delta$}
& \footnotesize{$0$} &
\footnotesize{$-1$} & \footnotesize{$1$} & \footnotesize{$0$} &  \footnotesize{$0$} & \footnotesize{$- \bar{\phi}_{\Delta \nu \lambda} \gamma^{\mu \nu \lambda}$}
\\
\hline
\hline 
\footnotesize{$C^{\star}$} & \footnotesize{$0$} & \footnotesize{$-2$} & $2$ & \footnotesize{$0$} & \footnotesize{$0$} & \footnotesize{$\nabla_\mu \nabla_\nu h^{\star \mu \nu} - \sigma \lambda^2 h^\star$}
\\
\hline
\footnotesize{$\bar{\tau}^{\star}_\Delta$} & \footnotesize{$1$} &
\footnotesize{$-2$} & $2$ & \footnotesize{$0$} &  \footnotesize{$0$} & \footnotesize{$- \nabla_\mu \bar{\phi}^{\star \mu}_\Delta + \frac{\lambda}{2} \bar{\phi}^{\star \mu}_\Omega \gamma_\mu M^\Omega{}_\Delta$} 
\\
\hline 
\end{tabular} 
\caption{Properties and BRST differentials of every field and antifield.}
\label{tableau PM unitary gauge}
\end{table}

\subsection{Cubic deformations of the gauge algebra}

As a starting point, we consider the list of all candidate deformations of the gauge algebra 
in the cohomology of $\gamma$ at pureghost number $2$, $H^{2}(\gamma)$, mixing the PM spin-2 
parameter with the spin-3/2 parameters. The only restriction is that we exclude candidates 
involving a $\gamma_5$ matrix that would break parity. There are only three 
independent terms:
\begin{equation}
\begin{aligned}
    a_2^{(1)} &= k^{(1)}_{\Delta \Sigma}\, \bar{\tau}^{\star \Delta} \gamma^\mu \tau^\Sigma \nabla_\mu C\;, \\
    a_2^{(2)} &= k^{(2)}_{\Delta \Sigma}\, \bar{\tau}^{\star \Delta} \tau^\Sigma C\;,  \\
    a_2^{(3)} &= k^{(3)}_{[\Delta \Sigma]}\, C^\star \bar{\tau}^{\Delta} \tau^\Sigma \;. 
\label{eq:PMsugraA2StartingPoint}    
\end{aligned}
\end{equation}

\subsection{Deformations of the gauge transformations}
The next step is to use the candidates \eqref{eq:PMsugraA2StartingPoint} as source for the equation \eqref{AdSDescentEqa1PMSUGRA}. We obtain a solution
\begin{equation}
\begin{aligned}
    a_1^{(1-2-3)}=\;&k_{\Delta \Sigma}^{(3)} \Bigl( 2\nabla_\mu h^{\star \mu \nu} \bar{\tau}^{\Delta} \phi_\nu^{\Sigma} + \lambda h^{\star \mu \nu } \bar{\phi}_\mu^{\Delta} \gamma_\nu \tau^{\Omega} {M}_{\Omega}{}^{\Sigma} + \lambda h^{\star \mu \nu } \bar{\tau}^\Delta \gamma_\mu \phi_\nu^{\Omega} {M}_{\Omega}{}^{\Sigma}\Bigr)\\
    &+k^{(1)}_{\Delta \Sigma} \Bigr( \bar{\phi}^{\star \Delta}_\mu \gamma^\nu \phi^{\Sigma \mu} \nabla_\nu C - \bar{\phi}^{\star \Delta}_\mu \gamma_\nu \tau^\Sigma h^{\mu\nu} + \lambda M_\Omega{}^\Sigma \,\bar{\phi}^{\star \Delta}_\mu \phi^{\Omega \mu} C\Bigr)\, 
    \label{eq:PMsugraA1IntermediateResult}
\end{aligned}
\end{equation}
under the constraints
\begin{equation}
\begin{aligned}
    \{k^{(1)},M\}&=0\,,\\
    \{k^{(3)},M\}&=0\,,\\
    k^{(2)}_{\Delta \Sigma}&=\lambda\, k^{(1)}_{\Delta \Omega}\, M_\Sigma{}^\Omega\;.
\end{aligned}
\end{equation}
This imposes that the matrix $M$ should take the form
\begin{equation}
    M=\pm \text{diag}(-1, \dots, -1, 1, \dots, 1)\;,
\end{equation}
with an equal number of $-1$'s and $+1$'s eigenvalues, 
with $\Delta = 1, \dots, 2 n$, providing the matrices $k^1$ and $k^3$ are 
anti-diagonal. Already, after solving this first equation, we see that the 
case $M_{\Delta \Sigma}=\pm\delta_{\Delta \Sigma}$ is excluded. In particular, 
in the case $n=1$ ($\Delta=1,2$), this means the R-symmetry is explicitly broken. 
A consequence of these constraints is also that the minimal number of spin-3/2 
fields is two, thereby excluding any non-Abelian coupling involving one PM 
spin-2 and one Majorana massless spin-3/2 field. This result is consistent with 
the result obtained in the previous section; it is not possible to render the 
supersymmetric transformations of \cite{Garcia-Saenz:2018wnw} local without 
introducing additional fields.  

\subsection{Cubic vertex}
We pursue by using the result \eqref{eq:PMsugraA1IntermediateResult} as source for the next equation in the descent \eqref{AdSDescentEqa0PMSUGRA}. In solving \eqref{AdSDescentEqa0PMSUGRA}, the solution $a_1$ to the previous equation \eqref{AdSDescentEqa1PMSUGRA} can always be combined with deformations $\bar{a}_1$ solutions to $\gamma \bar{a}_1=0$ that do not contribute to the gauge algebra. We find a solution 
\begin{equation}
    \delta \Bigl(  a_1^{(1-2-3)}+ \bar{a}_1\Bigr) + \gamma a^{\text{PM sugra}}_0 = \text{t.d.}\;,
\end{equation}
with
\begin{equation}
    \bar{a}_1 = b_{\Delta \Omega}\bar{\phi}^{\star \Delta \rho} \gamma^{\mu \nu} \tau^{\Omega} K_{\mu\nu \rho} + d_{\Delta \Omega} \bar{\phi}^{\star \Delta \nu} \phi_{\mu \nu}^\Omega \nabla^\mu C \;,
    \label{eq:PMSugraAddBarA_1}
\end{equation}
under the constraints  
\begin{equation}
\begin{aligned}
    b_{\Delta \Omega} =&\, b_{(\Delta \Omega)}\;, \qquad d_{\Delta \Omega}=-8 b_{\Delta \Omega}\;, \\
    k^{(1)}_{\Delta \Sigma} =&\, k^{(1)}_{[\Delta \Sigma]}=-2\lambda b_{(\Delta \Omega)} M_\Sigma{}^\Omega\; , \\
    b_{(\Delta \Omega)} =&\, \lambda k_{[\Delta \Sigma]}^{(3)} M_\Omega{}^\Sigma\; . \\
\end{aligned}
\end{equation}
The resulting interaction vertex is
\begin{equation}
\begin{aligned}
    a^{\text{PM sugra}}_0 = b_{\Delta \Omega} \Bigl\{ & -\frac{1}{2} \bar{\phi}^\Delta_\mu \gamma^{\mu\nu\rho}\gamma^{\alpha \beta} \phi_\rho^{\Omega} K_{\alpha \beta \nu}  \\
                                        & -2 \bar{\phi}^\Delta_\mu \gamma^{\nu \delta \rho} \phi_{\delta \rho}^{\Omega} h^\mu{}_\nu -4 \bar{\phi}^\Delta_\mu \gamma^{\mu \nu \rho} \phi_{\delta \rho}^{\Omega} h^{\nu \delta} +2 \bar{\phi}^\Delta_\mu \gamma^{\mu \nu \rho} \phi_{\nu \rho}^{\Omega} h \Bigr\}\;. \label{eq:PMSugraFinala0}
\end{aligned}
\end{equation}
The final deformation of the gauge transformations is 
\begin{equation}
\begin{aligned}
    a^{\text{PM sugra}}_1 = b_{\Delta \Gamma} &\Bigl\{ h^{\star\mu \nu} \Bigl( \nabla_\mu(-\frac{2}{\lambda}  \bar{\tau}^\Delta \phi_\nu^{\Sigma} M_\Sigma{}^\Gamma) + 2 \bar{\phi}_\nu^{\Delta} \gamma_\mu \tau^\Gamma \Bigr) \\
    &+\bar{\phi}_\nu^{\star \Delta} \Bigr( -2\lambda \gamma^\mu \phi^{\Sigma \nu} \nabla_\mu C M_\Sigma{}^\Gamma -2 \lambda^2 \phi^{\Gamma \nu} C +  8 \phi^{\Gamma \nu\mu}\nabla_\mu C\\
    &+ 2 \lambda \gamma^\mu \tau^\Sigma h^\nu{}_\mu M_{\Sigma}{}^\Gamma + \gamma^{\mu \delta} \tau^\Gamma K_{\mu \delta}{}^\nu \Bigl)  \Bigr\}\;, \label{eq:PMSugraFinala1}
\end{aligned}
\end{equation}
and the final deformation of the gauge algebra is 
\begin{equation}
    a^{\text{PM sugra}}_2 = -2\lambda b_{\Delta \Omega} M_{\Sigma}{}^\Omega \bar{\tau}^{\star \Delta} \gamma^\mu \tau^\Sigma \nabla_\mu C - 2 \lambda^2 b_{\Delta \Sigma} \bar{\tau}^{\star \Delta} \tau^\Sigma C + \frac{1}{\lambda} b_{\Delta \Omega} M_{\Sigma}{}^\Omega C^\star \bar{\tau}^\Delta \tau^\Sigma\;.
    \label{eq:PMSugraFinala2}
\end{equation}

\paragraph{Comparison with $\mathcal{N}_4=2$ supergravity}
Let us compare the PM supergravity cubic interacting model obtained in this 
section to the $\mathcal{N}_4=2$ supergravity model studied 
in \cite{Boulanger:2018fei} that shares the same spectrum of helicities 
$(\pm2,\pm3/2,\pm3/2,\pm1)$. In Subsection \ref{PM2kinematics}, we have presented 
the Stueckelberg formulation of the PM spin-2 field allowing to introduce a 
Stueckelberg gauge parameter $\varepsilon_\mu$ playing the role of linearised 
diffeomorphisms. In the unitary gauge where the Stueckelberg fields are set 
to zero, $\varepsilon_\mu$ is fixed to $\varepsilon_\mu = -\frac{\sigma}{2 \lambda} \nabla_\mu\pi $, 
where $\pi$ is the gauge parameter of the PM spin-2 field. 
In the following, we are going to compare the deformations of the gauge algebra, 
gauge transformations, and finally the interaction vertex of $\mathcal{N}_4=2$ 
supergravity obtained in \cite{Boulanger:2018fei} with the results of the present 
paper. In light of the Stueckelberg formulation of the PM spin-2 field, we identify 
$\nabla_\mu C$ in the partially massless case as a diffeomorphism ghost 
$\xi_\mu$ in $\mathcal{N}_4=2$ supergravity. 
The ghost field $C$  in the partially massless case is now replaced by 
the $U(1)$ ghost $\epsilon$ for the massless vector gauge field in 
$\mathcal{N}_4=2$ supergravity.  

We begin by the deformation of the gauge algebra \eqref{eq:PMSugraFinala2} that we 
compare to the corresponding gauge algebra of $\mathcal{N}_4=2$ supergravity given at 
Equation \eqref{Deformation:a2TrainaN2Susy}. The three algebra terms of the PM 
supergravity algebra \eqref{eq:PMSugraFinala2} are present in 
\eqref{Deformation:a2TrainaN2Susy}, but $a_2^{\text{sugra}}$ has two supplementary 
terms. In one term the diffeomorphism BV-antifield $\xi^\star_\mu$ appears, and in the 
other the diffeomorphism ghost appears through $\nabla_{[\mu} \xi_{\mu]}$. The first 
represents the commutator of two gauge supersymmetry transformations that gives a local 
diffeomorphism, and the second represents the commutator of a supersymmetry 
transformation and a local Lorentz transformation. The first term is absent because 
there is no way to represent $\xi^\star_\mu$ within the PM supergravity framework. The 
second term is absent because $\nabla_{[\mu} \xi_{\mu]}$ vanishes identically when 
$\xi^\star_\mu$ is interpreted as $\nabla_\mu C$. Therefore, we conclude that the PM 
supergravity gauge algebra \eqref{eq:PMSugraFinala2} is as close as possible to 
\eqref{Deformation:a2TrainaN2Susy}.

Let us rewrite the cubic results obtained in \cite{Boulanger:2018fei} for the gauge 
transformations of the $\mathcal{N}_4=2$ supergravity in AdS$_4$ with the same method: 
\begin{equation}
\begin{aligned}
    a_1^{\text{sugra}} = &\alpha_{3/2}\, k_{\Delta \Omega} \bigg[-h^{\star\mu\nu} \bar{\phi}_{\mu}^\Delta \gamma_\nu \tau^\Omega + \bar{\phi}^{\star \Delta\rho} \gamma_{\mu \nu} \phi_{\rho}^\Omega \nabla^{[\mu}\xi^{\nu]}  - \bar{\phi}^{\star\Delta\rho} \gamma^{\mu \nu} \tau^{\Omega} \nabla_{[\mu} h_{\nu]\rho} \\
    &+ \lambda \, \bar{\phi}^{\star\Delta\mu} \gamma^\nu \tau^{\Omega }h_{\mu \nu} -2 \lambda \bar{\phi}^{\star \Delta \mu} \gamma_\nu \phi_{\mu}^\Omega \xi^\nu -8 \bar{\phi}^{\star \Delta\mu} \phi_{\mu \nu}^\Omega \xi^\nu\bigg]\\
    &+y \,t_{\Delta \Omega} \bigg[ -2 A^{\star \mu} \bar{\phi}_\mu^\Delta \tau^\Omega -2 \lambda \bar{\phi}^{\star \Delta\mu} (\phi_{\mu}^\Omega \epsilon - \tau^\Omega A_{\mu}) - \frac{1}{2} \bar{\phi}^{\star\Delta}_{\mu} F_{\rho \sigma} \gamma^{\rho \sigma} \gamma^\mu \tau^\Omega\bigg]\;. \label{resultA1Sugra}
\end{aligned}  
\end{equation}
Here, the two spin-3/2 fields denoted $\phi_\mu^\Delta$ have a trivial mass matrix 
$M_{\Delta \Omega}= \pm \text{diag}(1,1)$, $h_{\mu \nu}$ stands for a massless spin-2 
field with diffeomorphism ghost $\xi_\mu$, and $A_\mu$ stands for a massless vector 
field with ghost $\epsilon$. In light of the Stueckelberg formulation of the PM spin-2 
field, we recall that we identify a $\xi_\mu$ in supergravity as a $\nabla_\mu C$ in 
the partially massless case. All terms in the first and second lines of 
\eqref{resultA1Sugra} are present in \eqref{eq:PMSugraFinala1}, except the term in 
$\nabla^{[\mu}\xi^{\nu]}$ that is identically zero in the partially massless case where 
$\xi_\mu$ is identified a $\nabla_\mu C$.
In the last line of \eqref{resultA1Sugra}, the only term in which the massless spin-1 
is not involved is present in \eqref{eq:PMSugraFinala1}. In fact, the only term present 
in the PM supergravity model and not in \eqref{resultA1Sugra} is the first term on the 
first line of \eqref{eq:PMSugraFinala1}. Notice that this term has the form 
$\nabla_{(\mu}\Tilde{\xi}_{\nu)}$ of a particular linearised diffeomorphism 
transformation of parameter $\Tilde{\xi}_\mu = -\frac{2}{\lambda} b_{\Delta \Gamma} \bar{\tau}^\Delta \phi_\mu^{\Sigma} M_\Sigma{}^\Gamma$.

We pursue by comparing the PM supergravity interaction vertex \eqref{eq:PMSugraFinala0} 
with the interaction vertex of $\mathcal{N}_4=2$ supergravity obtained in 
\cite{Boulanger:2018fei}. At first sight, the interaction vertices are the same, except 
for the helicity-1 modes that are enclosed in the PM spin-2 field in 
\eqref{eq:PMSugraFinala0}. The difference between the two is the symmetric matrix that 
contracts the $\Delta$ indices of the spin-3/2 fields. In the case of $\mathcal{N}_4=2$ 
supergravity in AdS$_4$, this matrix is proportional to the Kronecker delta 
$\delta_{\Delta \Omega}$, while in \eqref{eq:PMSugraFinala0} the matrix 
$b_{\Delta\Omega}$ is symmetric and anti-diagonal. This means that the partially 
massless supergravity vertex \eqref{eq:PMSugraFinala0} contains only cross terms 
of the type ``$\phi^1 \phi^2 h$'', while the supergravity vertex contains only 
diagonal terms of the type ``$\phi^1 \phi^1 h+\phi^2 \phi^2 h$''.

\subsection{Final result}

In this subsection, we rewrite and comment the obtained results without 
reference to the BRST-BV formalism. The Lagrangian at first order in 
deformation reads 
\begin{equation}
    L^{\text{PM sugra}} = L_0 + L_1\, ,
\end{equation}
with
\begin{equation}
    L_0 = -\frac{1}{4} K_{\mu \nu \rho} K^{\mu \nu \rho} + \frac{1}{2} K^\mu K_\mu -\frac{1}{2} \bar{\phi}_\mu^{\Delta} \gamma^{\mu \nu \rho} \nabla_\nu \phi_\rho^{\Sigma} \delta_{\Delta \Sigma} + \frac{\lambda}{2} \bar{\phi}_\mu^{\Delta} \gamma^{\mu \rho} \phi_\rho^{\Sigma} M_{\Delta \Sigma}\,,
    \label{PMSugraFinalResL0bis}
\end{equation}
and 
\begin{equation}
\begin{aligned}
    L_1 = b_{\Delta \Omega} \Bigl\{ & -\frac{1}{2} \bar{\phi}^\Delta_\mu \gamma^{\mu \nu \rho}\gamma^{\alpha \beta} \phi_\rho^{\Omega} K_{\alpha \beta \nu}  \\
                                        & -2 \bar{\phi}^\Delta_\mu \gamma^{\gamma\beta\rho} \phi_{\beta \rho}^{\Omega} h^\mu{}_\gamma -4 \bar{\phi}^\Delta_\beta \gamma^{\beta \gamma \rho} \phi_{\delta\rho}^{\Omega} h^{\delta}{}_\gamma +2 \bar{\phi}^\Delta_\beta \gamma^{\beta \gamma \rho} \phi_{\gamma \rho}^{\Omega} h \Bigr\}\,. \label{eq:PMSugraFinalResL1}
\end{aligned}
\end{equation}
We have shown that this interaction vertex is equivalent to the one obtained at 
equation (74) of \cite{Zinoviev:2024xta}.
At this order, this interacting model is invariant under gauge transformations
\begin{equation}
\begin{aligned}
    \delta h_{\alpha\beta} = &\nabla_\alpha \nabla_\beta \pi -\sigma \lambda^2\bar{g}_{\alpha\beta} \pi\\
                    & +b_{(\Delta \Gamma)}  \Bigl(\nabla_{(\alpha}(- \frac{2}{\lambda}  \bar{\phi}_{\beta)}^{\Delta} {\rho}^\Sigma  M_\Sigma{}^\Gamma )+ 2 \bar{\phi}_{(\alpha}^{\Delta} \gamma_{\beta)} \rho^\Gamma \Bigr)\,,
\end{aligned}
\end{equation}
\begin{equation}
\begin{aligned}
    \delta \phi^{\Delta \beta}=& \nabla^\beta \rho^\Delta + \frac{\lambda}{2} \gamma^\beta  \rho^\Sigma M_\Sigma{}^{\Delta}\\
    &+b^\Delta{}_\Gamma\Bigr( -2\lambda \gamma^\alpha \phi^{\Sigma \beta} \nabla_\alpha \pi M_\Sigma{}^\Gamma -2 \lambda^2 \phi^{\Gamma \beta} \pi +  8 \phi^{\Gamma  \beta \alpha}\nabla_\alpha \pi\\
    &+ 2 \lambda \gamma^\alpha \rho^\Sigma h^\beta{}_\alpha M_{\Sigma}{}^\Gamma + \gamma^{\alpha\gamma} \rho^\Gamma K_{\alpha\gamma}{}^\beta \Bigl)  \,.
\end{aligned}
\end{equation}
The gauge algebra takes the form
\begin{align}
    [\delta_{\Vec{\rho}}, \delta_{\Vec{\zeta}}] h_{\alpha \beta} &= \nabla_\alpha \nabla_\beta \Tilde{\pi} - \lambda^2 \bar{g}_{\alpha \beta} \Tilde{\pi}+ \mathcal{O}(b^2)\,,  \\
    [\delta_\pi, \delta_{\Vec{\rho}}] \phi_\alpha^\Delta &= \nabla_\alpha \Tilde{\rho}^\Delta + \frac{\lambda}{2} \gamma_\alpha  \Tilde{\rho}^\Sigma M_\Sigma{}^{\Delta} + \mathcal{O}(b^2)\,, \\
    [\delta_\pi, \delta_{\Vec{\rho}}] h_{\alpha \beta} &= \mathcal{O}(b^2)\,,\;\;\;\; [\delta_{\Vec{\rho}}, \delta_{\Vec{\zeta}}] \phi_\alpha^\Delta = \mathcal{O}(b^2)\,,   
\end{align}
with
\begin{align}
    \Tilde{\pi} & = \frac{2}{\lambda} \bar{\zeta}^\Delta \rho^\Sigma b_{\Delta\Gamma} M_\Sigma{}^\Gamma\,,\\
    \Tilde{\rho}^\Delta &= b^\Delta{}_{\Gamma}( 2\lambda   \gamma^\alpha \rho^\Sigma \nabla_\alpha \pi M_\Sigma{}^\Gamma+ 2 \lambda^2 \rho^\Gamma \pi)\,.
\end{align}
Rather suggestively, after a rescaling $b_{\Delta \Omega} \rightarrow -\frac{1}{8}b_{\Delta \Omega}$, the action with quadratic and cubic terms can be rewritten 
as the truncation up to (and including) the cubic order of
\begin{equation}
\begin{aligned}
    S^{\text{PM sugra}}= &\int  d^4x \,\bar{e}\,\Bigl(-\frac{1}{4} K_{\mu \nu \rho} K^{\mu \nu \rho} + \frac{1}{2} K^\mu K_\mu \Bigr)\\ +&\int  d^4x \,E_\Delta{}^\Omega (-\frac{1}{2} \bar{\phi}_\mu^\Delta \gamma^{abc} \Tilde{\phi}_{\nu \rho \Pi})(E^\mu_a){}_\Omega{}^\Theta (E^\nu_b){}_\Theta{}^\Gamma (E^{\rho}_c){}_\Gamma{}^\Pi \,,    
\end{aligned}
\end{equation}
with 
\begin{equation}
\begin{aligned}
    (G_{\mu \nu})_\Omega{}^\Delta &\equiv (E^a{}_\mu){}_\Omega{}^\Theta (E_{a\nu}){}_\Theta{}^\Delta = \bar{g}_{\mu \nu} \delta_\Omega{}^\Delta + h_{\mu \nu} b_{\Omega}{}^{\Delta}\;, \\
    (E^a{}_\mu){}_\Omega{}^\Delta &= \bar{e}^a{}_\mu \delta_\Omega{}^\Delta + \frac{1}{2} \bar{e}^{a\nu } h_{ \nu \mu} b_{\Omega}{}^\Delta\;,\\
    (E_a{}^\mu){}_\Omega{}^\Delta &= \bar{e}_a{}^\mu \delta_\Omega{}^\Delta - \frac{1}{2} \bar{e}_{a\nu} h^{\nu\mu }  b_{\Omega}{}^\Delta \;,\\
    E_{\Omega}{}^{\Delta} &= \bar{e}\delta_{\Omega}{}^{\Delta}  + \frac{1}{2}\bar{e} h b_{\Omega}{}^{\Delta} \;,\\
    \Tilde{\phi}_{\delta \rho}^\Delta &= \Tilde{\nabla}_{[\delta} \phi_{\rho]}^\Delta + \frac{\lambda}{2} \gamma_{[\delta} \phi_{\rho]}^\Omega M_\Omega{}^\Delta\;, \\
    \Tilde{\nabla}_{\delta} \phi_{\rho}^\Delta &= \nabla_{\delta} \phi_{\rho}^\Delta -\frac{1}{4}\nabla_{\alpha}h_{\beta \delta}\gamma^{\alpha \beta}\phi_{\rho}^\Omega b_{\Omega}{}^{\Delta}\;.
\end{aligned}
\end{equation}
This first order interacting model has the form of a gravitational minimal 
coupling, but where the role of the graviton is taken over by the partially 
massless spin-2 field. Compared to conventional minimal coupling in 
General Relativity, the partially massless spin-2 field is accompanied by 
the symmetric anti-diagonal matrix $b_{\Omega}{}^{\Delta}$ such that all 
the first order interactions terms are cross terms. 

\subsection{Global symmetries of the free model}
From the result of this first-order deformation computation, we can deduce global 
symmetries of the free Lagrangian. Consider global parameters $\Check{\pi}^\Delta$ and 
$\Check{\rho}$ solutions to $\delta_0 \phi^\Delta_\mu =0$ and $\delta_0 h_{\mu \nu}=0$:
\begin{align}
    \nabla^\beta \Check{\rho}^\Delta + \frac{\lambda}{2} \gamma^\beta  \Check{\rho}^\Sigma M_\Sigma{}^{\Delta}&=0\;,\label{GlobalParamSpin3/2}\\
    \nabla_\alpha \nabla_\beta \Check{\pi} - \lambda^2\bar{g}_{\alpha\beta} \Check{\pi} &=0\;.\label{GlobalParamSpin2}
\end{align}
Equation \eqref{GlobalParamSpin3/2} defines Killing spinors of the AdS$_4$ background. 
Moreover, if $\Check{\pi}$ is a solution to \eqref{GlobalParamSpin2}, then 
$\Check{\xi}_\mu=\nabla_\mu\Check{\pi}$ satisfies the conformal Killing equation 
\cite{Gibbons:1978ji, Deser:1983tm, Allen:1986ta}
\begin{equation}
    \nabla_\mu \Check{\xi}_\nu + \nabla_\nu \Check{\xi}_\mu = \frac{1}{2} \bar{g}_{\mu \nu}\nabla^\rho \Check{\xi}_\rho\;,
\end{equation}
and defines $\Check{\xi}_\mu$ as a conformal Killing vector of the AdS$_4$ background. 
Evaluating the first order gauge invariance equation
\begin{equation}
    \delta_0 L_1 + \delta_1 L_0 = 0\;,
\end{equation}
on global parameters $\Check{\pi}$ and $\Check{\rho}^\Delta$, 
we obtain $\Check{\delta}_1 L_0 = 0$, meaning that the first-order gauge 
transformations evaluated on global parameters define new symmetries of the free action 
with Lagrangian 
\eqref{PMSugraFinalResL0bis}. 
First, we obtain global fermionic transformations that mix the bosonic and fermionic 
sectors:
\begin{align}
    \Check{\delta}_{\Check{\rho}} h_{\alpha\beta} =&b_{\Delta \Gamma}  \Bigl(- \frac{2}{\lambda} \nabla_{(\alpha}( \bar{\phi}_{\beta)}^{\Delta} {\rho}^\Sigma  M_\Sigma{}^\Gamma )+ 2 \bar{\phi}_{(\alpha}^{\Delta} \gamma_{\beta)} \rho^\Gamma \Bigr)\;, \label{GlobalSusy32Tf}\\
    \Check{\delta}_{\Check{\rho}} \phi^{\Delta \beta}=&b^\Delta{}_\Gamma\Bigr(
    + 2 \lambda \gamma^\alpha \Check{\rho}^\Sigma h^\beta{}_\alpha M_{\Sigma}{}^\Gamma + \gamma^{\alpha\gamma} \Check{\rho}^\Gamma K_{\alpha\gamma}{}^\beta \Bigl) \;.
\end{align}
As expected, these transformations are very similar to the $\mathcal{N}_4=1$ 
supersymmetry transformations obtained in \cite{Garcia-Saenz:2018wnw} at equation 
(5.18) with a different spectrum. 
Second, we obtain a symmetry that mixes the spin-$3/2$ sector of 
\eqref{PMSugraFinalResL0bis} along a conformal Killing vector 
$\Check{\xi}_\mu=\nabla_\mu \Check{\pi}$. Up to a free gauge transformation 
$\Tilde{\delta}_0 \phi^{\Delta \beta}$ of free gauge parameter 
$\Tilde{\rho}^{\Delta}=b^\Delta{}_{\Gamma}\phi^{\Gamma \alpha}\Check{\xi}_\alpha$, this 
symmetry takes the form 
\begin{equation}
\Check{\delta}_{\Check{\pi}} \phi^{\Delta}_\mu = 
b^\Delta{}_\Gamma\Bigr(- \Check{\xi}^\nu\nabla_{\!\nu} \phi^{\Gamma}_{\mu} 
-\lambda\, \Check{\xi}^\nu\gamma_\nu \phi^{\Sigma}_\mu  M_{\Sigma}{}^{\Gamma} 
+ \lambda\, \gamma_\mu \Check{\xi}^\nu\phi^{\Sigma}_{\nu}  M_{\Sigma}{}^{\Gamma} 
- \tfrac{3}{2}\, \lambda^2 \,\phi^{\Gamma}_\mu  \Check{\pi}) 
+ \Tilde{\delta}_0 \phi^{\Delta}_{\mu}\;.
\label{GlobalConformallike32tf}
\end{equation}
This is very close to the recently derived conformal-like symmetry of the massless 
Dirac spin-3/2 field in dS$_4$ spacetime, see equation (6.21) of 
\cite{Letsios:2023awz}\footnote{The authors thank Vasileios Letsios for fruitful 
discussions on this subject.}. We leave the study of the commutators of these global 
symmetries for future work. 

\section{Obstruction to second order, a road to conformal supergravity}
\label{sec:conformalsugra}

At this stage, we have obtained a consistent cubic deformation $W_1^{\text{PM-sugra}}$ 
of the BV functional $W[\Phi^I,\Phi^*_I]= W_0 + W_1 + W_2 + \ldots$ 
up to first order in the coupling constants. The local functional we obtained, 
$W_1^{\text{PM-sugra}}=\int d^4x\, \sqrt{\bar{g}}\,(a_0 + a_1 + a_2)$, 
satisfies $(W_0,W_1) \equiv s W_1=0$ and is not $s$-exact in the space of 
local functionals.

To second order in the deformation, the BRST-BV master equation $(W,W)=0$ 
takes the form 
\begin{equation}
    s W_2 = -\frac{1}{2} (W_1, W_1)\,. 
\end{equation}
Once $W_2$ is expanded in antifield number, 
$W_2=\int d^4x\, \sqrt{\bar{g}}\, (b_0+ b_1+ b_2)$, one obtains the 
descent of equations
\begin{align}
    \delta b_1 + \gamma b_0 &= -   (a_1, a_0) + \nabla_\mu t^\mu_0\,, \\
    \delta b_2 + \gamma b_1 &= -\frac{1}{2} (a_1, a_1) -  (a_2, a_1) 
    + \nabla_\mu t^\mu_1 \,,\\
    \gamma b_2 &= -\frac{1}{2} (a_2, a_2) + \nabla_\mu t^\mu_2\,. \label{eq:MasterEqSecondOrderB2}
\end{align}
To solve this descent of equations for the triplet $(b_0, b_1,b_2)$, 
one starts with \eqref{eq:MasterEqSecondOrderB2} to be solved for $b_2$, 
using as a source the quantity $a_2$ that was obtained at previous order. 
In the particular case of the final $a_2$ of this section, 
see \eqref{eq:PMSugraFinala2}, the computation of the antibracket $(a_2, a_2)$ leads 
to obstructions ${\cal O}_i$ to the existence of $b_2$, in the form 
\begin{align}
    \mathcal{O}_1 &= 8 \lambda^2 b_{\Delta\Omega} M_{K}{}^{\Omega} b_{\Gamma}{}^K M_{\Theta}{}^{\Gamma} \bar{\tau}^{\star \Delta} \gamma^{\mu \nu} \tau^{\Theta} \nabla_\mu C \nabla_\nu C \;,\label{Obsterm1}\\
    \mathcal{O}_2 &= 16 \lambda^3 b_{\Delta \Omega} M_{\Pi}{}^{\Omega}b_\Sigma{}^\Pi   \bar{\tau}^{\star\Delta} \gamma^\mu \tau^\Sigma \nabla_\mu C C \;,\label{Obsterm2}\\
    \mathcal{O}_3 &= -8 b_{\Delta \Omega} M_\Pi{}^\Omega b_\Gamma{}^\Pi M_\Sigma{}^\Gamma C^\star \bar{\tau}^{(\Delta} \gamma^\mu \tau^{\Sigma)} \nabla_\mu C\;,\label{Obsterm3}\\
    \mathcal{O}_4 &= -4 \lambda b_{\Delta \Sigma} b_{\Gamma \Omega} M_{\Theta}{}^\Omega \bar{\tau}^{\star \Delta} \tau^\Sigma \bar{\tau}^\Gamma \tau^\Theta \;,\label{Obsterm4}\\
    \mathcal{O}_5 &= 4 \lambda b_{\Gamma \Theta} b_{\Delta \Omega} M_\Sigma{}^\Omega \bar{\tau}^{\star \Delta} \gamma^\mu \tau^\Sigma \bar{\tau}^\Gamma \gamma_\mu \tau^\Theta\;.\label{Obsterm5}
\end{align}
Unless one trivially sets the deformation structures $b_{\Delta \Omega}$ to zero, 
thereby setting $W_1$ to zero, there is no way to make these obstructions vanish.
This implies that the first-order deformation cannot be pushed to the next order 
in interactions. These obstructions signal a failure to the Jacobi identity 
for the gauge algebra to be satisfied.

The model considered so far contains the same spectrum of fields as the one of 
$\mathcal{N}_4=2$ pure supergravity in AdS$_4$, a theory that has been studied in 
\cite{Boulanger:2018fei} with the same cohomological techniques as the ones we 
have been following here. We recall that the on-shell spectrum of helicity modes 
of this model is $\{\pm2, \pm3/2, \pm3/2, \pm1\}\,$.
This is a remarkable point indeed, that motivated our investigations of 
the couplings between a PM spin-2 field with two massless gravitini around AdS$_4$.
We recall that the PM spin-2 field contains, in his decomposition in 
helicity modes in the flat limit $\lambda \rightarrow 0$, 
a massless spin-2 field as well as a massless spin-1 field, accounting for the 
helicity modes $(\pm2, \pm1)\,$. 

A noticeable difference between $\mathcal{N}_4=2$ pure sugra and the PM model 
we found up to cubic order is that, in the latter, the coupling between the 
spin-1 and the spin-$3/2$ modes is fixed by the coupling of the spin-2 modes
with the spin-$3/2$ modes, while in the perturbative reconstruction of 
$\mathcal{N}_4=2$ pure sugra around the free model around AdS$_4$, at 
cubic order one still has the freedom in the coupling constants between 
the gauge vector and the rest of the spectrum.  
In other words, all the coefficients in the vertex \eqref{eq:PMSugraFinalResL1} 
are already fixed at cubic-order, which means that the spin-1 modes are fixed 
to interact with the other modes, in a way that is not free to choose. 
By contrast, in $\mathcal{N}_4=2$ supergravity the coefficients in the vertices 
involving the spin-1 sector are fixed by the consistency relations imposed 
at second order. 

A possible way out to cure the obstructions \eqref{Obsterm1}-\eqref{Obsterm5}
we encountered in building a partially massless supergravity theory is the addition 
of a massless spin-2 field to the spectrum, a field that brings the diffeomorphism 
ghost $\xi^\mu$ into the BRST spectrum. 
The total deformation $a_2$ of the gauge algebra now reads
\begin{equation}
    a_2= a_2^{\text{PM Sugra}} + a_2^{\text{EH}} + a_2^{\text{Weyl}} + a_2^{\text{Sugra}}\;,
\end{equation}
where
\begin{align}
    a_2^{\text{EH}} &= \kappa\, \xi^{\star \mu} \xi^\nu \nabla_{[\mu} 
    \xi_{\nu]}\;,\label{Newa2EH}\\
    a_2^{\text{Weyl}} &= \alpha^{\text{Weyl}}\,( 2 
    \,C^\star \nabla^\mu C \,\xi_\mu + \lambda^2 
    \xi^{\star \mu} C \,\nabla_\mu C)\;,\label{Newa2Weyl}\\
    a_2^{\text{sugra}} &= \frac{1}{4}k_{\Delta \Omega}^{1}\xi^{\star \mu} 
    \bar{\tau}^\Delta \gamma_\mu \tau^\Omega + k_{\Delta \Omega}^{2}\bar{\tau}^{\star \Delta} 
    \gamma^{\mu \nu} \tau^\Omega \nabla_{[\mu} \xi_{\nu]}-2 \lambda 
    k_{\Delta\Omega}^{3}\bar{\tau}^{\star \Delta} \gamma^\mu \tau^{\Omega} \xi_\mu\;. 
    \label{Newa2Sugra}
\end{align}
The term $a_2^{\text{EH}}$ is the Einstein-Hilbert cubic deformation of the 
diffeomorphism algebra that uniquely leads to General Relativity \cite{Boulanger:2000rq}. 
The second deformation, $a_2^{\text{Weyl}}$, is the cubic deformation that mixes 
the gauge parameters of the massless and partially massless spin-2 fields. 
This deformation leads to conformal gravity at cubic order around the AdS$_4$ 
background \cite{Boulanger:2024hrb}. Indeed, the spectrum of conformal gravity linearised 
around (A)dS$_4$ spacetime is a massless spin-2 with a relatively ghostly partially 
massless spin-2 field \cite{Deser:1983tm}. 
In \cite{Boulanger:2024hrb}, it is proved that the relative sign 
and coefficients between $a_2^{\text{EH}}$ and $a_2^{\text{Weyl}}$ are fixed 
by the consistency conditions at second order. 
Finally, $a_2^{\text{sugra}}$ is a supergravity gauge algebra deformation 
that should lead to interactions between the massless spin-2 field 
and the spin-3/2 fields already present in the spectrum.
The candidate $a_2^{\text{sugra}}$ is based on the general form of the gauge
algebra deformation leading to $\mathcal{N}_4=2$ pure supergravity \cite{Boulanger:2018fei}. 
Considering two spin-3/2 fields with opposite signs of their respective mass-like terms, 
it remains to be shown if a deformation of the type $a_2^{\text{sugra}}$ is consistent. 

Certainly, if in \eqref{PMSugraFinalResL0bis} one considers four massless spin-3/2 
fields around AdS$_4$ with mass-like matrix 
$M_{\Delta \Omega}=\text{diag}(+1,+1,-1,-1)$, 
one can speculate that term $a_2^{\text{sugra}}$ is the direct sum of two deformations 
of the type $\mathcal{N}_4=2$ supergravity of \cite{Boulanger:2018fei}, 
which will require the introduction of extra spin-1 fields in the spectrum, 
for consistency at second order~\cite{Boulanger:2018fei}.

While the complete analysis has yet to be completed, we have shown that the 
addition of the 
algebraic deformations \eqref{Newa2EH}-\eqref{Newa2Sugra} produces new 
obstruction terms in 
$(a_2,a_2)$ that combine with \eqref{Obsterm1}-\eqref{Obsterm5} and 
potentially lead to a consistent model at the second order. 
These observations strongly indicate that there should be a complete, non-Abelian 
theory around AdS$_4$ that makes use of the vertices we found in this work, 
and whose spectrum of field is made of
\begin{itemize}
    \item one massless spin-2 field,
    \item one partially massless spin-2 field,
    \item two massless, real spin-3/2 fields,
    \item one massive, real spin-3/2 field, and
    \item one massless vector.
\end{itemize}
Remarkably, these fields make up the spectrum of ${\cal N}=1$ 
pure conformal supergravity expanded around AdS$_4$ \cite{Deser:1983tm,Fradkin:1985am}. 
From Section 4 and Appendix C of \cite{Bobev:2021oku} where 
${\cal N}=2$ conformal supergravity is considered, 
discarding the Yang-Mills, matter supermultiplets  
and extra $s_{\rm max}=3/2$ multiplets that appear once coupled 
to gravity when ${\cal N}>1$, see e.g. \cite{Freedman:2012zz} Sect. 6,
one can retrieve\footnote{From the Lagrangian and 
Tables 1 and 2 of \cite{Bobev:2021oku}, where the Einstein-Hilbert 
sector was added, one should formally take the limit $G_N\rightarrow \infty$ 
in their parameter 
$\delta = \frac{1}{2}\,+ \frac{1}{2}\,\sqrt{1+\frac{L^2}{8\pi G_N(c_1-c_2)}}$
so as to decouple the Einstein-Hilbert sector and only consider 
the higher-derivative terms (with coefficient $c_1-c_2$ therein) of 
pure conformal supergravity.}
the spectrum and conformal dimensions of the various fields of pure 
${\cal N}=1$ conformal supergravity expanded around AdS$_4$.
The result is summarized in Table \ref{confsugra} where we added 
in the last line the number of physical degrees of freedom of the 
corresponding fields. 
The spectrum of conformal (Weyl) gravity around AdS$_4$ had already been found in 
\cite{Deser:1983tm} while the fermionic sector of ${\cal N}=1$ 
pure conformal supergravity can be found in \cite{Fradkin:1985am,Bobev:2021oku}.
\begin{table}[t]
\centering
\begin{tabular}{|c||c|c|c|c|c|c|}
\hline
 $s$ & 2 & 2 & 3/2 & 3/2 & 3/2 & 1
\\
\hline\hline 
$\Delta$ & \footnotesize{3} & \footnotesize{2} & \footnotesize{5/2} & \footnotesize{5/2} & 
\footnotesize{3/2} & \footnotesize{2} \\ 
\hline 
$\#_{d.o.f.}$ & \footnotesize{2} & \footnotesize{4} & \footnotesize{2} & \footnotesize{2} & 
\footnotesize{4} & \footnotesize{2} 
\\ 
\hline
\end{tabular}
\caption{Set of fields of pure conformal supergravity expanded 
around AdS$_4$, with the conformal dimensions and number of physical degrees of freedom}
\label{confsugra}
\end{table}

Considering the gauge-fixed equation $(\Box - \lambda^2\,m^2)\varphi_s=0$
around AdS$_4\,$, where one has the mass formula 
$m_{\Delta,s}^2 = \Delta(\Delta - 3)-s$ for bosonic fields 
and $m_{\Delta,s}^2 = \Delta(\Delta - 3)-s-\frac{1}{4}$ for fermionic fields\footnote{A 
complete analysis in arbitrary dimension is given in 
\cite{Metsaev:1995re,Metsaev:1997nj,Metsaev:1998xg}, 
together with original references for the special case of $so(2,3)$ that 
we consider here. See also \cite{Basile:2016aen} for a review in both de Sitter and 
anti de Sitter.}, 
one can see that 
\begin{itemize}
    \item the field of spin-2 and conformal dimension 
$\Delta=3$ in Table \ref{confsugra} satisfies the linearised 
gauge-fixed field equation $(\Box + 2\lambda^2)\varphi_{\mu\nu}=0$.
This is the massless graviton with its two physical degrees of freedom;
  \item the field of spin-2 and conformal dimension 
$\Delta=2$ in Table \ref{confsugra} satisfies the linearised 
gauge-fixed field equation $(\Box + 4\,\lambda^2)h_{\mu\nu}=0$.
This is the partially massless spin-2 field with its four physical degrees 
of freedom; 
   \item the two fields of spin-3/2 and conformal dimension 
$\Delta=5/2$ in Table \ref{confsugra} satisfy the linearised 
gauge-fixed field equation $(\Box + 3\,\lambda^2)\phi_{\mu}=0$.
These are  two massless gravitini with their two physical degrees of freedom, 
where the field equation is recovered from \eqref{Dalembert32} with 
$m^2=0$; 
   \item the field of spin-3/2 and conformal dimension 
$\Delta=3/2$ in Table \ref{confsugra} satisfies the linearised field equation $(\Box + 4\,\lambda^2)\psi_\mu=0$
which coincides with the  equation \eqref{Dalembert32} 
when $\omega=0$, the massive fields $\psi_\mu$ that enter the vertex 
\eqref{InteractingTheoryVertices}. 
This is a massive gravitino with its four physical degrees of freedom
with helicities $(\pm 3/2,\pm1/2)$ in the flat limit;
  \item finally, the field of spin-1 and conformal dimension 
$\Delta=2$ in Table \ref{confsugra} satisfies the linearised 
gauge-fixed field equation $(\Box + 3\,\lambda^2)A_{\mu}=0$.
This is a massless vector field with its two physical degrees of freedom.
\end{itemize}
It was found in \cite{Deser:1983tm} that the first two spin-2 fields in the 
list above make up the spectrum of Weyl (conformal) gravity around AdS$_4$.

The computations \eqref{Obsterm1}-\eqref{Obsterm5} on the obstruction to the Jacobi identity 
for the gauge algebra strongly indicate that the only consistent 
non-Abelian theory for the interactions of a partially massless 
spin-2 field with massive and massless spin-3/2 fields around AdS$_4$ 
is ${\cal N}=1$ conformal supergravity, indeed a non-unitary theory
at the classical level.
Our results indicate that the mass-like terms for the two 
massless gravitini of ${\cal N}=1$ conformal supergravity expanded around 
AdS$_4$ must have opposite signs.

\section{Conclusions}
\label{sec:conclusions}

In this paper, we have made a detailed study of the possible interactions 
between a partially massless spin-2 field and spin-3/2 fields, 
massless and massive. The motivation behind our work is partly
based on the paper \cite{Garcia-Saenz:2018wnw} where a minimal 
partially massless supermultiplet around AdS$_4$ was found that 
contains the three types of fields cited above, plus a gauge vector.
The findings of \cite{Garcia-Saenz:2018wnw} stimulates the search for 
a partially massless supergravity model that would gauge the rigid 
supersymmetry carried by the supermultiplet. 
Our results show that the presence of the massless spin-3/2 field 
of the PM supermultiplet is not enough to localise the rigid supersymmetry: 
there is no way to make local the global supersymmetry algebra represented 
on the minimal PM supermultiplet of \cite{Garcia-Saenz:2018wnw} without 
introducing extra fields. 
On the way to this result, we found two vertices that couple the massive 
and massless spin-3/2 fields to the gauge vector of the PM supermultiplet, 
in a way that deforms the Abelian gauge transformations of the free theory, 
but not the gauge algebra that remains Abelian.

In order to search for a non-Abelian model where the PM spin-2 field 
would not remain sterile, we took advantage of 
the observation that the spectrum of ${\cal N}=2$ pure supergravity 
around AdS$_4$ is identical to the set of fields 
$\{h_{\mu\nu},\phi^\Delta_\mu\}$, $\Delta = 1, 2$,  
where $h_{\mu\nu}$ is the partially massless spin-2 field and 
$\{\phi^\Delta_\mu\}$, $\Delta = 1, 2$ a doublet of massless gravitini, 
and classified all the possible non-Abelian deformations of the gauge 
algebra which can lead to a deformation of the Lagrangian.
We found the existence of a single non-Abelian vertex, that
coincides with the vertex recently presented in \cite{Zinoviev:2024xta},
in a different formalism and using different field representations. 
In the representation we use where the two massless gravitini  
are Majorana spinors, we showed a very suggestive analogy between this vertex 
and the minimal coupling in ${\cal N}=2$ pure supergravity, except 
that now, the role of the graviton is taken over by the partially massless 
spin-2 field. This vertex is a candidate for a partially massless 
supergravity. 

We pushed our analysis to second order in deformation, at the level of 
the Jacobi identity for the gauge algebra, and showed that the  
non-Abelian partially massless supergravity vertex is obstructed.
We then added a massless graviton to the spectrum with its corresponding 
diffeomorphism gauge parameter, and argued that this addition is potentially 
sufficient to cure the obstruction to the Jacobi identity of the resulting
gauge algebra, that now includes diffeomorphisms. 
Remarkably, the final spectrum of fields coincides with 
the spectrum of ${\cal N}=1$ pure conformal supergravity around AdS$_4$, 
therefore viewed 
as the only consistent (although non-unitary at the classical level) and 
non-Abelian theory for what one could call partially massless supergravity. 

In this work, as long as there was at least one massless Majorana spin-3/2 field 
in the spectrum, the background considered was AdS$_4$ with its 
negative cosmological constant. On the other hand, the results we obtained 
on the coupling of massive spin-3/2 fields with a vector gauge fields were 
valid in both AdS$_4$ and dS$_4$ backgrounds. 
The relations we made with conformal supergravity stimulates 
us to consider an expansion of that theory around dS$_4$, using the recent 
findings and spinor field representations of \cite{Letsios:2023awz,Letsios:2023tuc}.
In fact, it is probably not a coincidence that we recuperated the rigid conformal 
transformation of these latter works, as our findings suggest that the closure 
of the transformations laws \eqref{GlobalConformallike32tf}, very close to those 
found in \cite{Letsios:2023awz,Letsios:2023tuc}, leads to a representation of the 
superconformal algebra. We hope to pursue our investigations along those lines 
in the future, and investigate to which extend and with which field  
representation for the spinor fields one can expand ${\cal N}=1$ 
conformal supergravity around dS$_4$.

\acknowledgments
It is a pleasure to thank Thomas Basile, Nicolay Bobev, Guillaume Bossard, 
William Delplanque, Marc Henneaux, Kurt Hinterbichler and Vasileios Letsios 
for discussions. 
We are particularly grateful to Yurii M. Zinoviev for many discussions and 
for sharing his draft of paper \cite{Zinoviev:2024xta}, prior to its submission 
to the arxives. The work of S.T. was supported by the FNRS ASP fellowship FC 54793 
MassHighSpin. The work of N.B. was partly supported by the FNRS PDR grant 
T.0047.24.



\providecommand{\href}[2]{#2}\begingroup\raggedright\endgroup


\begin{thebibliography}{10}

\bibitem{Deser:1983tm}
S.~Deser and R.I.~Nepomechie, \emph{{Anomalous Propagation of Gauge Fields in Conformally Flat Spaces}}, \href{https://doi.org/10.1016/0370-2693(83)90317-9}{\emph{Phys. Lett. B} {\bfseries 132} (1983) 321}.

\bibitem{Deser:1983mm}
S.~Deser and R.I.~Nepomechie, \emph{{Gauge Invariance Versus Masslessness in De Sitter Space}}, \href{https://doi.org/10.1016/0003-4916(84)90156-8}{\emph{Annals Phys.} {\bfseries 154} (1984) 396}.

\bibitem{Higuchi:1986py}
A.~Higuchi, \emph{{Forbidden Mass Range for Spin-2 Field Theory in De Sitter Space-time}}, \href{https://doi.org/10.1016/0550-3213(87)90691-2}{\emph{Nucl. Phys. B} {\bfseries 282} (1987) 397}.

\bibitem{Higuchi:1986wu}
A.~Higuchi, \emph{{Symmetric Tensor Spherical Harmonics on the $N$ Sphere and Their Application to the De Sitter Group SO($N$,1)}}, \href{https://doi.org/10.1063/1.527513}{\emph{J. Math. Phys.} {\bfseries 28} (1987) 1553}.

\bibitem{Deser:2001pe}
S.~Deser and A.~Waldron, \emph{{Gauge invariances and phases of massive higher spins in (A)dS}}, \href{https://doi.org/10.1103/PhysRevLett.87.031601}{\emph{Phys. Rev. Lett.} {\bfseries 87} (2001) 031601} [\href{https://arxiv.org/abs/hep-th/0102166}{{\ttfamily hep-th/0102166}}].

\bibitem{Deser:2001us}
S.~Deser and A.~Waldron, \emph{{Partial masslessness of higher spins in (A)dS}}, \href{https://doi.org/10.1016/S0550-3213(01)00212-7}{\emph{Nucl. Phys. B} {\bfseries 607} (2001) 577} [\href{https://arxiv.org/abs/hep-th/0103198}{{\ttfamily hep-th/0103198}}].

\bibitem{Deser:2001wx}
S.~Deser and A.~Waldron, \emph{{Stability of massive cosmological gravitons}}, \href{https://doi.org/10.1016/S0370-2693(01)00523-8}{\emph{Phys. Lett. B} {\bfseries 508} (2001) 347} [\href{https://arxiv.org/abs/hep-th/0103255}{{\ttfamily hep-th/0103255}}].

\bibitem{Baumann:2017jvh}
D.~Baumann, G.~Goon, H.~Lee and G.L.~Pimentel, \emph{{Partially Massless Fields During Inflation}}, \href{https://doi.org/10.1007/JHEP04(2018)140}{\emph{JHEP} {\bfseries 04} (2018) 140} [\href{https://arxiv.org/abs/1712.06624}{{\ttfamily 1712.06624}}].

\bibitem{Franciolini:2017ktv}
G.~Franciolini, A.~Kehagias and A.~Riotto, \emph{{Imprints of Spinning Particles on Primordial Cosmological Perturbations}}, \href{https://doi.org/10.1088/1475-7516/2018/02/023}{\emph{JCAP} {\bfseries 02} (2018) 023} [\href{https://arxiv.org/abs/1712.06626}{{\ttfamily 1712.06626}}].

\bibitem{Goon:2018fyu}
G.~Goon, K.~Hinterbichler, A.~Joyce and M.~Trodden, \emph{{Shapes of gravity: Tensor non-Gaussianity and massive spin-2 fields}}, \href{https://doi.org/10.1007/JHEP10(2019)182}{\emph{JHEP} {\bfseries 10} (2019) 182} [\href{https://arxiv.org/abs/1812.07571}{{\ttfamily 1812.07571}}].

\bibitem{SupernovaCosmologyProject:1998vns}
{\scshape Supernova Cosmology Project} collaboration, \emph{{Measurements of $\Omega$ and $\Lambda$ from 42 High Redshift Supernovae}}, \href{https://doi.org/10.1086/307221}{\emph{Astrophys. J.} {\bfseries 517} (1999) 565} [\href{https://arxiv.org/abs/astro-ph/9812133}{{\ttfamily astro-ph/9812133}}].

\bibitem{SupernovaSearchTeam:1998fmf}
{\scshape Supernova Search Team} collaboration, \emph{{Observational evidence from supernovae for an accelerating universe and a cosmological constant}}, \href{https://doi.org/10.1086/300499}{\emph{Astron. J.} {\bfseries 116} (1998) 1009} [\href{https://arxiv.org/abs/astro-ph/9805201}{{\ttfamily astro-ph/9805201}}].

\bibitem{Planck:2018vyg}
{\scshape Planck} collaboration, \emph{{Planck 2018 results. VI. Cosmological parameters}}, \href{https://doi.org/10.1051/0004-6361/201833910}{\emph{Astron. Astrophys.} {\bfseries 641} (2020) A6} [\href{https://arxiv.org/abs/1807.06209}{{\ttfamily 1807.06209}}].

\bibitem{LIGOScientific:2018dkp}
{\scshape LIGO Scientific, Virgo} collaboration, \emph{{Tests of General Relativity with GW170817}}, \href{https://doi.org/10.1103/PhysRevLett.123.011102}{\emph{Phys. Rev. Lett.} {\bfseries 123} (2019) 011102} [\href{https://arxiv.org/abs/1811.00364}{{\ttfamily 1811.00364}}].

\bibitem{LIGOScientific:2020tif}
{\scshape LIGO Scientific, Virgo} collaboration, \emph{{Tests of general relativity with binary black holes from the second LIGO-Virgo gravitational-wave transient catalog}}, \href{https://doi.org/10.1103/PhysRevD.103.122002}{\emph{Phys. Rev. D} {\bfseries 103} (2021) 122002} [\href{https://arxiv.org/abs/2010.14529}{{\ttfamily 2010.14529}}].

\bibitem{LIGOScientific:2021sio}
{\scshape LIGO Scientific, Virgo, KAGRA} collaboration, \emph{{Tests of General Relativity with GWTC-3}},  \href{https://arxiv.org/abs/2112.06861}{{\ttfamily 2112.06861}}.

\bibitem{Zinoviev:2008ze}
Y.M.~Zinoviev, \emph{{Frame-like gauge invariant formulation for massive high spin particles}}, \href{https://doi.org/10.1016/j.nuclphysb.2008.09.020}{\emph{Nucl. Phys. B} {\bfseries 808} (2009) 185} [\href{https://arxiv.org/abs/0808.1778}{{\ttfamily 0808.1778}}].

\bibitem{Boulanger:2018shp}
N.~Boulanger, A.~Campoleoni and I.~Cortese, \emph{{Dual actions for massless, partially-massless and massive gravitons in (A)dS}}, \href{https://doi.org/10.1016/j.physletb.2018.05.046}{\emph{Phys. Lett. B} {\bfseries 782} (2018) 285} [\href{https://arxiv.org/abs/1804.05588}{{\ttfamily 1804.05588}}].

\bibitem{Boulanger:2019zic}
N.~Boulanger, C.~Deffayet, S.~Garcia-Saenz and L.~Traina, \emph{{Theory for multiple partially massless spin-2 fields}}, \href{https://doi.org/10.1103/PhysRevD.100.101701}{\emph{Phys. Rev. D} {\bfseries 100} (2019) 101701} [\href{https://arxiv.org/abs/1906.03868}{{\ttfamily 1906.03868}}].

\bibitem{Joung:2014aba}
E.~Joung, W.~Li and M.~Taronna, \emph{{No-Go Theorems for Unitary and Interacting Partially Massless Spin-Two Fields}}, \href{https://doi.org/10.1103/PhysRevLett.113.091101}{\emph{Phys. Rev. Lett.} {\bfseries 113} (2014) 091101} [\href{https://arxiv.org/abs/1406.2335}{{\ttfamily 1406.2335}}].

\bibitem{Joung:2019wwf}
E.~Joung, K.~Mkrtchyan and G.~Poghosyan, \emph{{Looking for partially-massless gravity}}, \href{https://doi.org/10.1007/JHEP07(2019)116}{\emph{JHEP} {\bfseries 07} (2019) 116} [\href{https://arxiv.org/abs/1904.05915}{{\ttfamily 1904.05915}}].

\bibitem{Boulanger:2024hrb}
N.~Boulanger, S.~Garcia-Saenz, S.~Pan and L.~Traina, \emph{{Cubic interactions for massless and partially massless spin-1 and spin-2 fields}}, \href{https://doi.org/10.1007/JHEP11(2024)019}{\emph{JHEP} {\bfseries 11} (2024) 019} [\href{https://arxiv.org/abs/2407.05865}{{\ttfamily 2407.05865}}].

\bibitem{Boulanger:2023lgd}
N.~Boulanger, G.~Lhost and S.~Thom\'ee, \emph{{Consistent Couplings between a Massive Spin-3/2 Field and a Partially Massless Spin-2 Field}}, \href{https://doi.org/10.3390/universe9110482}{\emph{Universe} {\bfseries 9} (2023) 482} [\href{https://arxiv.org/abs/2310.05522}{{\ttfamily 2310.05522}}].

\bibitem{Garcia-Saenz:2018wnw}
S.~Garcia-Saenz, K.~Hinterbichler and R.A.~Rosen, \emph{{Supersymmetric Partially Massless Fields and Non-Unitary Superconformal Representations}}, \href{https://doi.org/10.1007/JHEP11(2018)166}{\emph{JHEP} {\bfseries 11} (2018) 166} [\href{https://arxiv.org/abs/1810.01881}{{\ttfamily 1810.01881}}].

\bibitem{Fradkin:1985am}
E.S.~Fradkin and A.A.~Tseytlin, \emph{{Conformal Supergravity}}, \href{https://doi.org/10.1016/0370-1573(85)90138-3}{\emph{Phys. Rept.} {\bfseries 119} (1985) 233}.

\bibitem{Bobev:2021oku}
N.~Bobev, A.M.~Charles, K.~Hristov and V.~Reys, \emph{{Higher-derivative supergravity, AdS$_{4}$ holography, and black holes}}, \href{https://doi.org/10.1007/JHEP08(2021)173}{\emph{JHEP} {\bfseries 08} (2021) 173} [\href{https://arxiv.org/abs/2106.04581}{{\ttfamily 2106.04581}}].

\bibitem{Becchi:1975nq}
C.~Becchi, A.~Rouet and R.~Stora, \emph{{Renormalization of Gauge Theories}}, \href{https://doi.org/10.1016/0003-4916(76)90156-1}{\emph{Annals Phys.} {\bfseries 98} (1976) 287}.

\bibitem{Tyutin:1975qk}
I.V.~Tyutin, \emph{{Gauge Invariance in Field Theory and Statistical Physics in Operator Formalism}}, {\emph{Lebedev preprint} {\bfseries 39} (1975) } [\href{https://arxiv.org/abs/0812.0580}{{\ttfamily 0812.0580}}].

\bibitem{Batalin:1981jr}
I.A.~Batalin and G.A.~Vilkovisky, \emph{{Gauge Algebra and Quantization}}, \href{https://doi.org/10.1016/0370-2693(81)90205-7}{\emph{Phys. Lett. B} {\bfseries 102} (1981) 27}.

\bibitem{Batalin:1983ggl}
I.A.~Batalin and G.A.~Vilkovisky, \emph{{Quantization of Gauge Theories with Linearly Dependent Generators}}, \href{https://doi.org/10.1103/PhysRevD.28.2567}{\emph{Phys. Rev. D} {\bfseries 28} (1983) 2567}.

\bibitem{BatalinErratum}
I.A.~Batalin and G.A.~Vilkovisky, \emph{Erratum: Quantization of gauge theories with linearly dependent generators}, \href{https://doi.org/10.1103/PhysRevD.30.508}{\emph{Phys. Rev. D} {\bfseries 30} (1984) 508}.

\bibitem{Stueckelberg:1957zz}
E.C.G.~Stueckelberg, \emph{{Th\'eorie de la radiation de photons de masse arbitrairement petite}}, {\emph{Helv. Phys. Acta} {\bfseries 30} (1957) 209}.

\bibitem{Ruegg:2003ps}
H.~Ruegg and M.~Ruiz-Altaba, \emph{{The Stueckelberg field}}, \href{https://doi.org/10.1142/S0217751X04019755}{\emph{Int. J. Mod. Phys. A} {\bfseries 19} (2004) 3265} [\href{https://arxiv.org/abs/hep-th/0304245}{{\ttfamily hep-th/0304245}}].

\bibitem{Boulanger:2018dau}
N.~Boulanger, C.~Deffayet, S.~Garcia-Saenz and L.~Traina, \emph{{Consistent deformations of free massive field theories in the Stueckelberg formulation}}, \href{https://doi.org/10.1007/JHEP07(2018)021}{\emph{JHEP} {\bfseries 07} (2018) 021} [\href{https://arxiv.org/abs/1806.04695}{{\ttfamily 1806.04695}}].

\bibitem{Boulanger:2018fei}
N.~Boulanger, B.~Julia and L.~Traina, \emph{{Uniqueness of $ \mathcal{N} $ = 2 and 3 pure supergravities in 4D}}, \href{https://doi.org/10.1007/JHEP04(2018)097}{\emph{JHEP} {\bfseries 04} (2018) 097} [\href{https://arxiv.org/abs/1802.02966}{{\ttfamily 1802.02966}}].

\bibitem{Barnich:1993vg}
G.~Barnich and M.~Henneaux, \emph{{Consistent couplings between fields with a gauge freedom and deformations of the master equation}}, \href{https://doi.org/10.1016/0370-2693(93)90544-R}{\emph{Phys. Lett. B} {\bfseries 311} (1993) 123} [\href{https://arxiv.org/abs/hep-th/9304057}{{\ttfamily hep-th/9304057}}].

\bibitem{Henneaux:1997bm}
M.~Henneaux, \emph{{Consistent interactions between gauge fields: The Cohomological approach}}, \href{https://doi.org/10.1090/conm/219/03070}{\emph{Contemp. Math.} {\bfseries 219} (1998) 93} [\href{https://arxiv.org/abs/hep-th/9712226}{{\ttfamily hep-th/9712226}}].

\bibitem{Zinoviev:2024xta}
Y.M.~Zinoviev, \emph{{Partially massless spin 2 and supersymmetry}},  \href{https://arxiv.org/abs/2412.04982}{{\ttfamily 2412.04982}}.

\bibitem{Boulanger:2000rq}
N.~Boulanger, T.~Damour, L.~Gualtieri and M.~Henneaux, \emph{{Inconsistency of interacting, multigraviton theories}}, \href{https://doi.org/10.1016/S0550-3213(00)00718-5}{\emph{Nucl. Phys. B} {\bfseries 597} (2001) 127} [\href{https://arxiv.org/abs/hep-th/0007220}{{\ttfamily hep-th/0007220}}].

\bibitem{Boulanger:2018adg}
N.~Boulanger, A.~Campoleoni, I.~Cortese and L.~Traina, \emph{{Spin-2 twisted duality in (A)dS}}, \href{https://doi.org/10.3389/fphy.2018.00129}{\emph{Front. in Phys.} {\bfseries 6} (2018) 129} [\href{https://arxiv.org/abs/1807.04524}{{\ttfamily 1807.04524}}].

\bibitem{Pilch:1984aw}
K.~Pilch, P.~van Nieuwenhuizen and M.F.~Sohnius, \emph{{De Sitter Superalgebras and Supergravity}}, \href{https://doi.org/10.1007/BF01211046}{\emph{Commun. Math. Phys.} {\bfseries 98} (1985) 105}.

\bibitem{Townsend:1977qa}
P.K.~Townsend, \emph{{Cosmological Constant in Supergravity}}, \href{https://doi.org/10.1103/PhysRevD.15.2802}{\emph{Phys. Rev. D} {\bfseries 15} (1977) 2802}.

\bibitem{Fronsdal:1978rb}
C.~Fronsdal, \emph{{Massless Fields with Integer Spin}}, \href{https://doi.org/10.1103/PhysRevD.18.3624}{\emph{Phys. Rev. D} {\bfseries 18} (1978) 3624}.

\bibitem{Fronsdal:1978vb}
C.~Fronsdal, \emph{{Singletons and Massless, Integral Spin Fields on de Sitter Space (Elementary Particles in a Curved Space. 7.}}, \href{https://doi.org/10.1103/PhysRevD.20.848}{\emph{Phys. Rev. D} {\bfseries 20} (1979) 848}.

\bibitem{Fang:1979hq}
J.~Fang and C.~Fronsdal, \emph{{Massless, Half Integer Spin Fields in De Sitter Space}}, \href{https://doi.org/10.1103/PhysRevD.22.1361}{\emph{Phys. Rev. D} {\bfseries 22} (1980) 1361}.

\bibitem{Flato:1978qz}
M.~Flato and C.~Fronsdal, \emph{{One Massless Particle Equals Two Dirac Singletons: Elementary Particles in a Curved Space. 6.}}, \href{https://doi.org/10.1007/BF00400170}{\emph{Lett. Math. Phys.} {\bfseries 2} (1978) 421}.

\bibitem{Fronsdal:1975eq}
C.~Fronsdal and R.B.~Haugen, \emph{{Elementary Particles in a Curved Space. 3}}, \href{https://doi.org/10.1103/PhysRevD.12.3810}{\emph{Phys. Rev. D} {\bfseries 12} (1975) 3810}.

\bibitem{Breitenlohner:1982jf}
P.~Breitenlohner and D.Z.~Freedman, \emph{{Stability in Gauged Extended Supergravity}}, \href{https://doi.org/10.1016/0003-4916(82)90116-6}{\emph{Annals Phys.} {\bfseries 144} (1982) 249}.

\bibitem{Henneaux:1994lbw}
M.~Henneaux and C.~Teitelboim, \emph{{Quantization of Gauge Systems}}, Princeton University Press (8, 1994).

\bibitem{Bittermann:2020xkl}
N.~Bittermann, S.~Garcia-Saenz, K.~Hinterbichler and R.A.~Rosen, \emph{{$ \mathcal{N} $ = 2 supersymmetric partially massless fields and other exotic non-unitary superconformal representations}}, \href{https://doi.org/10.1007/JHEP08(2021)115}{\emph{JHEP} {\bfseries 08} (2021) 115} [\href{https://arxiv.org/abs/2011.05994}{{\ttfamily 2011.05994}}].

\bibitem{Velo:1969bt}
G.~Velo and D.~Zwanziger, \emph{{Propagation and quantization of Rarita-Schwinger waves in an external electromagnetic potential}}, \href{https://doi.org/10.1103/PhysRev.186.1337}{\emph{Phys. Rev.} {\bfseries 186} (1969) 1337}.

\bibitem{Johnson:1960vt}
K.~Johnson and E.C.G.~Sudarshan, \emph{{Inconsistency of the local field theory of charged spin 3/2 particles}}, \href{https://doi.org/10.1016/0003-4916(61)90030-6}{\emph{Annals Phys.} {\bfseries 13} (1961) 126}.

\bibitem{Deser:2000dz}
S.~Deser, V.~Pascalutsa and A.~Waldron, \emph{{Massive spin 3/2 electrodynamics}}, \href{https://doi.org/10.1103/PhysRevD.62.105031}{\emph{Phys. Rev. D} {\bfseries 62} (2000) 105031} [\href{https://arxiv.org/abs/hep-th/0003011}{{\ttfamily hep-th/0003011}}].

\bibitem{Deser:2001dt}
S.~Deser and A.~Waldron, \emph{{Inconsistencies of massive charged gravitating higher spins}}, \href{https://doi.org/10.1016/S0550-3213(02)00199-2}{\emph{Nucl. Phys. B} {\bfseries 631} (2002) 369} [\href{https://arxiv.org/abs/hep-th/0112182}{{\ttfamily hep-th/0112182}}].

\bibitem{Zinoviev:2006im}
Y.M.~Zinoviev, \emph{{On massive spin 2 interactions}}, \href{https://doi.org/10.1016/j.nuclphysb.2007.02.005}{\emph{Nucl. Phys. B} {\bfseries 770} (2007) 83} [\href{https://arxiv.org/abs/hep-th/0609170}{{\ttfamily hep-th/0609170}}].

\bibitem{Gibbons:1978ji}
G.W.~Gibbons and M.J.~Perry, \emph{{Quantizing Gravitational Instantons}}, \href{https://doi.org/10.1016/0550-3213(78)90434-0}{\emph{Nucl. Phys. B} {\bfseries 146} (1978) 90}.

\bibitem{Allen:1986ta}
B.~Allen, \emph{{The Graviton Propagator in De Sitter Space}}, \href{https://doi.org/10.1103/PhysRevD.34.3670}{\emph{Phys. Rev. D} {\bfseries 34} (1986) 3670}.

\bibitem{Letsios:2023awz}
V.A.~Letsios, \emph{{New conformal-like symmetry of strictly massless fermions in four-dimensional de Sitter space}}, \href{https://doi.org/10.1007/JHEP05(2024)078}{\emph{JHEP} {\bfseries 05} (2024) 078} [\href{https://arxiv.org/abs/2310.01702}{{\ttfamily 2310.01702}}].

\bibitem{Freedman:2012zz}
D.Z.~Freedman and A.~Van~Proeyen, \emph{{Supergravity}}, Cambridge Univ. Press, Cambridge, UK (5, 2012), \href{https://doi.org/10.1017/CBO9781139026833}{10.1017/CBO9781139026833}.

\bibitem{Metsaev:1995re}
R.R.~Metsaev, \emph{{Massless mixed symmetry bosonic free fields in d-dimensional anti-de Sitter space-time}}, \href{https://doi.org/10.1016/0370-2693(95)00563-Z}{\emph{Phys. Lett. B} {\bfseries 354} (1995) 78}.

\bibitem{Metsaev:1997nj}
R.R.~Metsaev, \emph{{Arbitrary spin massless bosonic fields in d-dimensional anti-de Sitter space}}, \href{https://doi.org/10.1007/BFb0104614}{\emph{Lect. Notes Phys.} {\bfseries 524} (1999) 331} [\href{https://arxiv.org/abs/hep-th/9810231}{{\ttfamily hep-th/9810231}}].

\bibitem{Metsaev:1998xg}
R.R.~Metsaev, \emph{{Fermionic fields in the d-dimensional anti-de Sitter space-time}}, \href{https://doi.org/10.1016/S0370-2693(97)01446-9}{\emph{Phys. Lett. B} {\bfseries 419} (1998) 49} [\href{https://arxiv.org/abs/hep-th/9802097}{{\ttfamily hep-th/9802097}}].

\bibitem{Basile:2016aen}
T.~Basile, X.~Bekaert and N.~Boulanger, \emph{{Mixed-symmetry fields in de Sitter space: a group theoretical glance}}, \href{https://doi.org/10.1007/JHEP05(2017)081}{\emph{JHEP} {\bfseries 05} (2017) 081} [\href{https://arxiv.org/abs/1612.08166}{{\ttfamily 1612.08166}}].

\bibitem{Letsios:2023tuc}
V.A.~Letsios, \emph{{Unconventional conformal invariance of maximal depth partially massless fields on dS$_{4}$ and its relation to complex partially massless SUSY}}, \href{https://doi.org/10.1007/JHEP08(2024)147}{\emph{JHEP} {\bfseries 08} (2024) 147} [\href{https://arxiv.org/abs/2311.10060}{{\ttfamily 2311.10060}}].

\end{thebibliography}


\end{document}